\documentclass[fleqn,usenatbib,useAMS]{mnras}

\usepackage{amssymb}

\usepackage{newtxtext,newtxmath}

\usepackage[T1]{fontenc}
\usepackage{ae,aecompl}

\usepackage{graphicx}	
\usepackage{amsmath}	
\usepackage{amssymb}	
\usepackage{color}

\newcommand{\eqb}{\begin{equation}}
\newcommand{\eqe}{\end{equation}}

\def\mpcoh{{\,h^{-1}\,\rm Mpc}}
\def\hompc{\,h\,{\rm Mpc}^{-1}}
\def\citejap#1{\citeauthor{#1}\ \citeyear{#1}}

\def\m@th{\mathsurround=0pt }
\def\eqalign#1{\null\,\vcenter{\openup1\jot \m@th
 \ialign{\strut\hfil$\displaystyle{##}$&$\displaystyle{{}##}$\hfil
 \crcr#1\crcr}}\,}
\def\xib{\overline{\xi}}
\def\xibb{\overline{\overline{\xi}}}

\title[Modelling RSD in galaxy-group cross-correlations]{Galaxy and Mass Assembly (GAMA): Probing galaxy-group correlations in redshift space with the halo streaming model}

\author[Q. Hang et al.]{
Qianjun Hang$^{1,2}$\thanks{E-mail: qhang@roe.ac.uk},
John A. Peacock$^{1}$,
Shadab Alam$^{1}$,
Yan-Chuan Cai$^{1}$,
Katarina Kraljic$^{1,3}$,
\newauthor
Marcel van Daalen$^4$,
M. Bilicki$^5$,
B.W. Holwerda$^6$, J.Loveday$^7$
\\
$^{1}$Institute for Astronomy, University of Edinburgh, Royal Observatory Edinburgh, Blackford Hill, Edinburgh EH9 3HJ, UK\\
$^{2}$Department of Physics and Astronomy, University College London, Gower Street, London WC1E 6BT, UK\\
$^{3}$Aix Marseille Universit\'e, CNRS, CNES, UMR 7326, Laboratoire d'Astrophysique de Marseille, Marseille, France\\
$^4$Leiden Observatory, Leiden University, P.O. Box 9513, 2300 RA Leiden, The Netherlands \\
$^5$Center for Theoretical Physics, Polish Academy of Sciences, al. Lotnik\'ow 32/46, 02-668 Warsaw, Poland  \\
$^6$Department of Physics and Astronomy, University of Louisville, Natural Science Building 102, 40292 KY Louisville, USA \\
$^7$Astronomy Centre, University of Sussex, Falmer, Brighton BN1 9QH, UK
}

\pubyear{2022}

\begin{document}
\label{firstpage}
\pagerange{\pageref{firstpage}--\pageref{lastpage}}
\maketitle

\begin{abstract}
We have studied the galaxy-group cross-correlations in redshift space for the Galaxy And Mass Assembly (GAMA) Survey. We use a set of mock GAMA galaxy and group catalogues to develop and test a novel `halo streaming' model for redshift-space distortions. This treats 2-halo correlations via the streaming model, plus an empirical 1-halo term derived from the mocks, allowing accurate modelling into the nonlinear regime. In order to probe the robustness of the growth rate inferred from redshift-space distortions, we divide galaxies by colour, and divide groups according to their total stellar mass, calibrated to total mass via gravitational lensing. 
We fit our model to correlation data, to obtain estimates of the perturbation growth rate, $f\sigma_8$, validating parameter errors via the dispersion between different mock realizations.
In both mocks and real data, we demonstrate that the results are closely consistent between different subsets of the group and galaxy populations, considering the use of correlation data down to some minimum projected radius, $r_{\rm min}$.  For the mock data, 
we can use the halo streaming  model to below $r_{\rm min} = 5\mpcoh$, finding that
all subsets yield growth rates within about 3\% of each other, and consistent with the true value.  For the actual GAMA data, the results are limited by cosmic variance: $f\sigma_8=0.29\pm 0.10$ at an effective redshift of 0.20; but there is every reason to expect that this method will yield precise constraints from larger datasets of the same type, such as the DESI bright galaxy survey.
\end{abstract}




\section{Introduction}

The large-scale structure in the galaxy distribution has a long history of providing cosmological information. The first constituents of the inhomogeneous galaxy density field to be identified were the rich clusters, which today we see as marking the sites of exceptionally massive haloes of dark matter. Proceeding down the halo mass spectrum, we find progressively less rich groups of galaxies, leading to systems dominated by a single $L_*$ galaxy, such as the Local Group (e.g. \citejap{2018ARA&A..56..435W}). All these systems have been familiar constituents of the Universe since the first telescopic explorations of the sky, but it took rather longer to appreciate that they were connected as part of the cosmic web of voids \& filaments (see e.g. \citejap{2016:japweb} for some selective history). In part, the history here showed a complex interaction of theory and observation, since redshift surveys through the 1980s lacked the depth and sampling to reveal the cosmic web with complete clarity. For a period, it was therefore a question of asking whether the real Universe displayed the same structures that were predicted in numerical simulations of structure formation in the Cold Dark Matter model (\citejap{1996:bondweb}). But since those times, there has been an increasing confidence that galaxy groups are indeed particularly extreme nonlinear points in the general field of cosmic density fluctuations, and this makes them interesting in two ways. First of all, groups are readily identified in galaxy surveys, providing a relatively robust dataset \citep{2pigg, jr:robotham}. Secondly, their nonlinear nature makes them an informative probe of theory. Modelling nonlinear behaviour is by its nature challenging compared to linear theory, but by studying structure formation further into the nonlinear regime, we have the chance to test the robustness of our cosmological conclusions.

Our specific aim in this direction is to use galaxy groups as a probe of the cosmological peculiar velocity field. Such deviations from uniform expansion must exist through continuity, and density concentrations such as groups should be associated with an average infall velocity in regions surrounding the groups. The amplitude of these velocities depends in part on the strength of gravity on cosmological scales, and the peculiar velocity field has thus increasingly been seen as a means of probing the nature of gravity and testing alternative theories (e.g. \citejap{Jain2010J}). Although it is possible to probe peculiar velocities directly using absolute distance indicators \citep{Davis2011}, the most powerful tool has been Redshift Space Distortions (RSD). These arise inevitably in the study of the 3D galaxy distribution because the distances to galaxies observed on the sky are inferred from their redshifts, $z$, via the standard relation
\eqb
d(z)=\int_0^z\frac{c\,dz'}{H(z')},
\eqe
where $c$ is the speed of light and $H(z)=\dot{a}/a$ is the Hubble parameter. But this equation does not give the true distances, because Doppler shifts from the peculiar velocities modify the observed redshift: $1+z\rightarrow (1+z)(1+v_r/c)$, where $v_r$ is the radial component of the peculiar velocity. If we then use the observed redshift as if it were a true indicator of distance, we obtain a distribution of galaxies in `redshift space' -- in which the apparent properties of galaxy clustering are distorted in an anisotropic way. 

These distortions have characteristics that depends on scale:
outside large density concentrations, galaxies fall coherently together under gravity; while the orbital velocities inside dark-matter haloes are effectively randomised. The latter effect convolves the redshift-space density field in the radial direction, leading to the characteristic radial elongations of high-density regions known as `Fingers of God' (FoG).
RSD due to coherent flows in the linear regime were first studied by \citet{jr:kaiser1987}. The growth factor $f$ is defined by
\eqb
f\equiv\frac{\partial \ln\delta}{\partial \ln a}\simeq \Omega_m(z)^{0.55},
\label{eq: f}
\eqe
where $\delta$ is the matter overdensity, $a$ is the expansion factor, and $\Omega_m$ is the matter fraction; 
The approximation for $f(\Omega_m)$ only applies for flat $\Lambda$CDM models in standard gravity \citep{jr:lahav1991, 1998ApJ...508..483W, 2005PhRvD..72d3529L}. 
In Fourier space, and in the small-angle limit of a distant observer, 
the matter power spectra in redshift space and in real space are related by
\eqb
P^s_m(k,\mu)=P^r_m(k)\,(1+f\mu^2)^2,
\label{eq: kaiser}
\eqe
where $\mu$ is the cosine of the angle between the wave-vector $k$ and the line of sight. 
This simple equation was highly influential from its first appearance \citep{jr:kaiser1987}, as it offered the
chance of measuring $\Omega_m$ from measuring the RSD anisotropy. But eventually goals shifted
as $\Omega_m$ became very well determined from other routes (especially the Cosmic Microwave Background).
Following \cite{jr:guzzo2008}, the modern view is therefore
to emphasise that the growth rate for a given density is also proportional
to the strength of gravity, so that RSD can be used as a test of theories of gravity.

The RSD signal has been measured by a number of surveys, including the 2dFGRS at $z\simeq 0.2$ \citep{jr:peacock2001,jr:hawkins}; the 6dFGS at $z\simeq 0.1$ \citep{2012MNRAS.423.3430B}; the SDSS BOSS \& eBOSS surveys at $z\simeq 0.6$ \citep{jr:reid2012,jr:alam2017,Alam2021}; and at $z\simeq 1$ by the 8m VVDS and VIPERS surveys \citep{jr:guzzo2008, vipers2017}. For the GAMA survey at $z\simeq 0.4$, aspects of RSD were studied by \cite{jr:blake2013} and \citet{jr:loveday}, who measured the pair-wise velocity dispersion to small scales and as a function of luminosity. The above studies all focused on galaxy auto-correlations.

The challenge in modelling RSD is that truly linear modes are rare. In observation, large scales are affected by cosmic variance due to the finite survey volume. \citet{jr:mcdonaldseljak2009} proposed the use of multiple tracers in order to overcome cosmic variance, although in practice the improvement is slight \citep{jr:blake2013}. To gain more information, one needs to probe smaller scales, where the effect of non-linearity can systematically bias the results \citep{jr:delatorre2012}. 

One possible solution to this dilemma is to use galaxy groups to probe the velocity field.
Due to the small random virial velocity of the central galaxy at the group centre, the coherent large-scale infall velocities of groups are dominant down to intermediate and small scales. The group auto-correlation would thus have reduced FoG, aiding the extraction of the linear growth rate \citep{2001ApJ...554..873P,jr:mohammad}. In practice, the group catalogue in GAMA is sparse, with a number density of $4.3\times10^{-3}h^3{\rm Mpc}^{-3}$ between $0.1<z<0.3$, and measurements of the auto-correlation will have high statistical noise. The cross-correlation between groups and galaxies is thus an intermediate route, which effectively improves the statistical power while still reducing the non-linear pairwise velocities at small scales. The clustering of GAMA groups has been recently studied in \citet{2021MNRAS.506...21R}, and the present work extends this study to further subsets of the data, concentrating in more detail on their different RSD signals.

Our aim here is thus to test the robustness of RSD methods down to small or intermediate scales using multiple tracers involving galaxy groups. By cross-correlating galaxies of different colours, and groups in different mass bins, we examine the consistency of the inferred cosmological results between the subsamples. In order to pursue this investigation, we develop a new model for RSD in cross-correlation, involving a combination of the halo model and the streaming model, which we implement by including some information taken from mock data. Throughout the analysis, we adopt the WMAP7 Cosmology \citep{2011ApJS..192...18K} with $\sigma_8=0.81$, $\Omega_m=0.27$, $h=0.70$, and $n_s=0.967$, consistent with the mock catalogue.

The GAMA data set and its mocks are detailed in Section~\ref{sec:GAMA-data} followed by Section~\ref{sec:measurement-statistics} where we introduce the statistics for measuring the 2-point function in the data.  In Section~\ref{sec:measurements} we present the resulting 2D correlation function measurements for sub-samples, In Section~\ref{sec:models} we discuss the theoretical modelling of RSD in galaxy-group cross-correlations, and in section~\ref{sec:model-fitting} we confront this modelling with real and mock GAMA data. The models are validated in Section~\ref{sec:results-mocks} via detailed comparison with the GAMA mocks, where we establish the scales to which the different theories can work without bias; we present the  fitting of the real GAMA data in Section~\ref{sec:results-gama}. Finally, we summarize the work in Section~\ref{sec:conclusions}.

\newpage
\section{GAMA data and mocks}\label{sec:GAMA-data}

This analysis is based on the Galaxy And Mass Assembly (GAMA) spectroscopic survey.
This was conducted using the 2dF facility at the Anglo-Australian 4m telescope over 210 nights between 2008 \& 2014, accumulating spectra of 265\,958 distinct galaxies.
Together with existing data, this yielded a catalogue of 330\,542 redshifts over five survey fields totalling 250\,deg$^2$, with a mean redshift of $z\simeq0.2$ \citep{Driver_2022}. The three main fields near the equator, G09, G12, and G15 are used here, each covering an area of $12\times 5 \,\deg ^2$. The survey has an extinction-corrected $r$-band flux limit of $r<19.8$, based on SDSS photometry. 

The overall redshift completeness of the GAMA equatorial region is 98.5\%: this high completeness was achieved by a large number of repeated visits to 2dF fields covering the survey area in different ways. This property is greatly advantageous for small scale galaxy and group studies compared to much larger surveys such as BOSS, where fibre collisions can lead to substantial undercounting of close galaxy pairs and thus bias the measured galaxy 2-point correlation function \citep{2012ApJ...756..127G}.

The present analysis uses the DR3 data release \citep{gamaDR3}, which differs slightly from the final data release, DR4 \citep{Driver_2022}. DR4 implements revised flux completeness limits through the use of new KiDS photometry. The original SDSS limit of $r<19.8$ was trimmed in DR4 to $r<19.58$ for 98\% completeness in the equatorial fields. 
Galaxies for the present study are selected from the \texttt{SpecObjv27} DMU (Data Management Unit), with CMB frame redshifts adopted from \texttt{DistancesFramesv14}. We apply the following criteria: redshift quality \texttt{nQ} $\geq3$, angular completeness mask $>80\%$, and visual classification \texttt{VIS\_CLASS} $=0,1,255$\footnote{\texttt{VIS\_CLASS} $=0$: Not visually inspected but suspicious based on SDSS flags; \texttt{VIS\_CLASS} $=1$: Visually inspected and a valid target; \texttt{VIS\_CLASS} $=255$: Not visually inspected but should be OK based on SDSS flags.}. 
The spectroscopic redshifts are computed by the code \texttt{runz} \citep{2011MNRAS.413..971D}, which has a $1\sigma$ redshift error of 50\,km\,s$^{-1}$ in terms of peculiar velocity \citep{jr:gamaDR2}.
In order to compute correlation statistics, it is essential to accompany the galaxy sample with a knowledge of the survey selection in angle and redshift. As usual, this information is captured by a random catalogue \texttt{Randomsv02} of fictitious unclustered galaxies; this catalogue was generated by \cite{jr:farrow2015} from the actual GAMA galaxy catalogue using a modified method following \cite{2011MNRAS.416..739C}. The idea of this method is to clone each galaxy $n$ times and distribute them randomly within the maximum volume $V_{\rm max}$ that the galaxy can be observed given the survey magnitude limits,
\begin{equation}
n=n_{\rm clones}\frac{V_{\rm max}}{V_{\rm max, dc}},
\end{equation}
where $n_{\rm clones}=400$ is the total number of randoms divided by data, and $V_{\rm max, dc}$ is the maximum volume weighted by overdensity $\Delta(z)$. This method is iterated until $\Delta(z)$ converges, and the redshift distribution of the resultant random catalogue is smooth without large scale features (see Fig.~4 in \citejap{jr:farrow2015}).

The official GAMA group catalogue (G3C) was constructed by \citet{jr:robotham}. Most of the groups are found within $z\lesssim0.35$ (see Fig.~16 in \citejap{jr:robotham}): thus we impose a redshift cut $0.1<z<0.3$ for the groups. 
The group catalogue is derived using an anisotropic friends-of-friends (FoF) algorithm calibrated against an $N$-body mock catalogue. However, in order to have consistently defined groups in the GAMA mocks (see Section~\ref{sec:mocks}), we do not use the official G3C catalogue. Instead, we apply a similar FoF group finder algorithm due to \citet{jr:katarina1} to both data and mocks (see \citejap{jr:katarina2} for an application to GAMA). The main difference between the two algorithms is the parameterisation of the linking length, and a detailed description of the algorithm and assessment of the group reconstruction quality can be found in the Appendix of \cite{jr:katarina1}.

In addition to the above selections, we further split galaxies and groups into subsamples based on galaxy colour and group mass. The number of selected galaxies and groups in each GAMA field and for each subsample is summarised in Table~\ref{tab:selected galaxies and groups}. We describe the selection in more detail below. 

\begin{table}
	\centering
	\caption{Number of selected galaxies and groups from GAMA fields with redshifts $0.1<z<0.3$ and flux limit $r<19.8$. Galaxies are split into two colour classes, red and blue, by Eq.~\ref{eq:g-i}. Groups are split into three stellar mass bins: 40\% (LM), 50\% (MM), and 10\% (HM) by mass ranking from low to high, covering the mass range $\log_{10}(M_*/h^{-2}M_{\odot})=9.5-12.5$.}
	\label{tab:selected galaxies and groups}
	\begin{tabular}{llrrr}
		\hline
		Number of & & G09 & G12 & G15\\
		\hline
		Galaxies& Blue& 17\,335 & 18\,719 & 19\,053\\
		        &Red & 20\,584 & 22\,155 & 21\,141\\
        &{\bf Total}& 37\,919& 40\,874& 40\,194\\
		\hline
        Groups&LM& 1877 & 2084 & 2054 \\
        &MM& 2347& 2606& 2569\\
        &HM& 470& 522& 514\\
        &{\bf Total} & 4694& 5212& 5137\\
		\hline
		G3C & {\bf Total} & 4937 & 5367 & 5358\\
		\hline
	\end{tabular}
\end{table}

\subsection{GAMA galaxy colour selection}
\label{sec:GAMA galaxies and groups} 

\begin{figure}
\centering
	\includegraphics[width=0.47\textwidth]{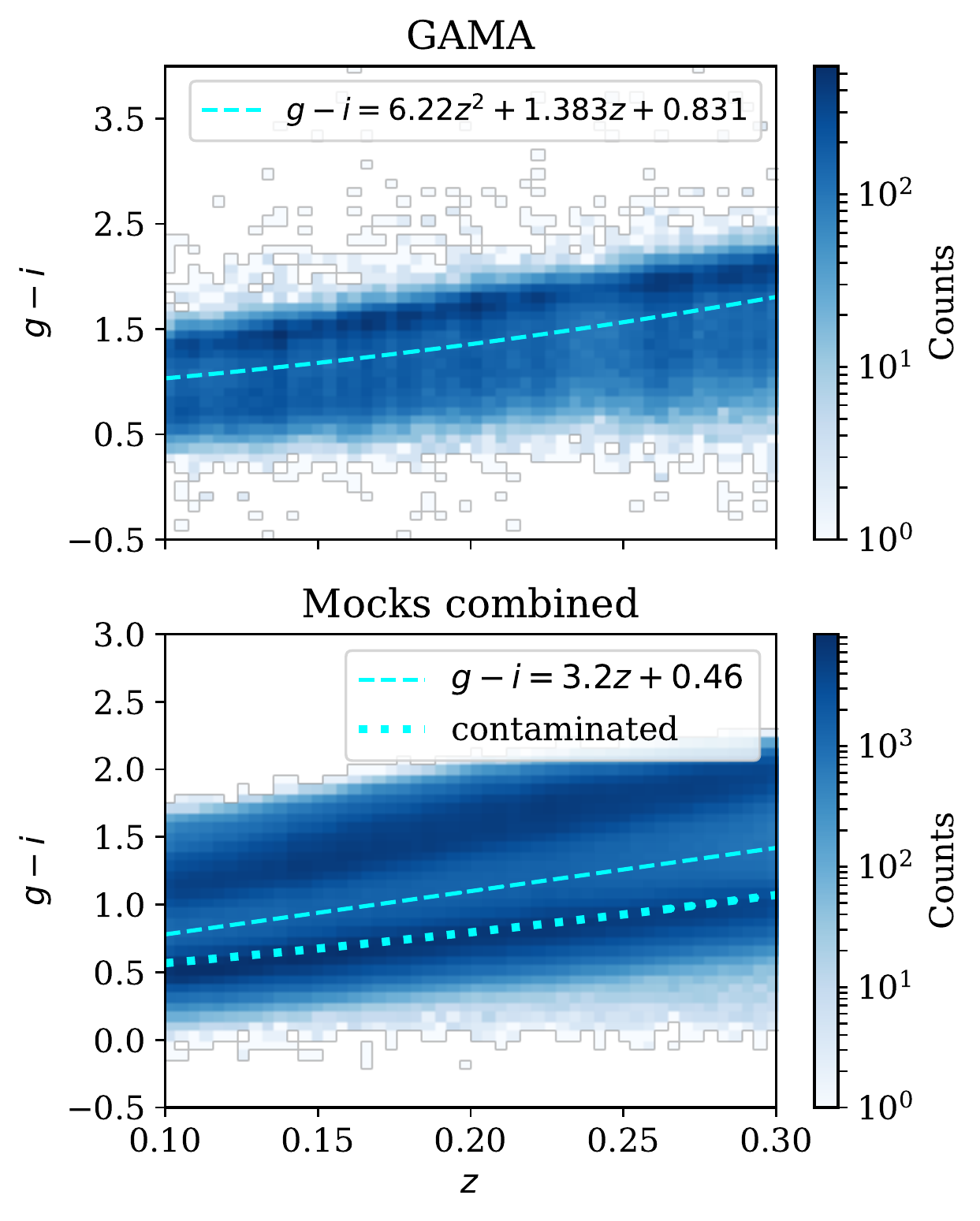}
    \caption{Distribution of the $g-i$ colour of galaxies in the redshift range $0.1<z<0.3$, for the real GAMA data (upper panel) and the average of 25 mocks combined (lower panel). Red and blue populations are separated by the dashed cyan lines. The cut in GAMA is chosen such that both GAMA and mocks have similar red and blue fractions at any given redshift. The dotted lines show the cut for an alternative `contaminated' red sample (see text).}
    \label{fig: color_split_gama_mock}
\end{figure}

Galaxies are divided into two populations that are known to have distinct clustering properties: the `red' galaxies, which tend to be older, with little or no active star formation, and the `blue' cloud, where galaxies are younger with active star formation. 
To obtain the galaxy colours, we use the extinction corrected SDSS magnitudes from the \texttt{TilingCatv46} DMU. It is however non-trivial to separate the galaxy population into these subsets, because the colour distribution is continuous without gaps: elaborate approaches have been discussed in e.g. \cite{jr:taylor2015}.
For the purpose of this study, we adopt a simple quadratic cut in the apparent $g-i$ colour versus redshift plane:
\begin{equation}
g-i = 6.220z^2+1.383z+0.831.
\label{eq:g-i}
\end{equation}
A cut of this form is motivated empirically by the apparent bimodality in the colour-redshift plane, as shown in Fig.~\ref{fig: color_split_gama_mock}.
The precise location of the cut was adjusted in order to match the red and blue fraction at each redshift in the GAMA data with the corresponding result in the mocks (which are discussed below in Section~\ref{sec:mocks}).  The overall fraction of red or blue galaxies is very close to 0.5, and it changes only slightly with redshift: at the low redshift end, the red and blue fractions are similar, while towards higher redshifts, the fraction of red galaxies increases mildly until $z\sim0.2$, and the difference in the red and blue fraction then becomes small at $z\sim0.3$. We create random catalogues for the red and blue galaxy subsamples where required by applying this smoothly-varying colour balance to the redshift distribution of the main random catalogue.

\subsection{GAMA group mass selection}

Groups are accepted with $\geq 2$ group members, and the centre of the group is determined by the most (more) massive member in terms of stellar mass. The 2-member systems make up $66\%$ of the total groups in the GAMA data, but are likely to have poor fidelity. Thus, we emphasise that having the same group finder algorithm for the data and mocks is vital in order for these low-fidelity groups to be comparable. There are several approaches for determining the group centre. The simplest choice is to select the most massive member to be the central galaxy, and assume that it overlaps with the halo centre. Other approaches include determining a weighted centre by averaging over the positions of the group members, or iteratively excluding members that are most distantly separated (see e.g. \citejap{jr:robotham}). The iterative centres are used in the G3C catalogue, and it is shown in \cite{jr:robotham} that the agreement with using the brightest group galaxy (BCG) as group centre is $95\%$ for groups with $N\geq 5$, and that both BCG and iterative centres give highly consistent results for $2\leq N\leq 4$ compared with the mock, and that the BCG centres are only degraded by about $3\%$ compared to the iterative centres. The effects of different group centre choices on the group-galaxy cross-correlation concern mainly the 1-halo regime at $r\leq 1\mpcoh$, and the correlation functions converge on larger scales \citep{2005MNRAS.362..711Y}.

The halo mass of GAMA groups was found to be tightly correlated with the group total luminosity \citep[][]{jr:han, Viola2015, 2022MNRAS.510.5408R}, based on using stacked weak lensing measurements to determine the mass distribution of the GAMA groups.
The halo mass of groups is related to the $r$-band luminosity $L_{\rm grp}$ via
\begin{equation} 
M_{h} = M_p\left(\frac{L_{\rm grp}}{L_0}\right)^\alpha,
\label{eq: mh vs L}
\end{equation}
where 
$L_0=2\times10^{11}h^{-2}L_{\odot}$, $\log_{10}(M_p/h^{-1}M_{\odot})=13.48-0.08\pm0.12$, and $\alpha=1.08+0.01\pm0.22$ \citep[][]{jr:han}. The group halo mass can be used to compute the expected mean group bias, as shown in Section~\ref{sec: group bias}, where we also consider alternative mass calibrations.
It should be noted that in these works, only groups with three members or above are used. \citet{jr:robotham} showed that the mass function is noisier, but not biased when including two-member groups. Although these systems are individually of low reliability, this aspect should be allowed for by the mock catalogue, allowing us to gain the statistical advantage of using a larger sample.
The luminosity is computed from the apparent $r$-band magnitude:
\begin{equation}
-2.5\log(L/L_{\odot})=m-K(z)-5\log(d_L)-25-M_\odot,
\end{equation}
where $K(z)$ the $k$-correction up to $z=0$ (\texttt{kcorr\_z00}), $d_L$ is the luminosity distance, and $M_\odot=4.67$ is the $r$-band absolute magnitude of the sun. The luminosity distance is expressed with unit $\mpcoh$ so that the luminosity has units of $h^{-2}L_\odot$.

The total luminosity of the group is computed in \cite{jr:robotham} via
\begin{equation}
L_{\rm FoF}=B L_{\rm ob}\frac{\int_{-30}^{-14}10^{-0.4M_r}\phi_{\rm GAMA}(M_r)\,dM_r}{\int_{-30}^{M_{r-{\rm lim}}}10^{-0.4M_r}\phi_{\rm GAMA}(M_r)\,dM_r},
\end{equation}
where $L_{\rm ob}$ is the total observed luminosity in the $r_{\rm AB}$ band, $B=1.04$ is the correction for median unbiased estimate for $N\geq5$ groups, and $M_{r-{\rm lim}}$ is the absolute magnitude limit of the group depending on the redshift $z$. $\phi_{\rm GAMA}$ is the luminosity function defined in \cite{jr:robotham}. 
The luminosity function at the faint end for GAMA galaxies is well approximated by $\phi\propto L^{-1}\exp(-L/L_*)$ \citep{2012MNRAS.420.1239L}. Thus in practice we take the simpler approach of estimating a total group luminosity by scaling the observed luminosity by a redshift-dependent correction factor $\exp(z^2/z_*^2)$ with $z_*=0.33$, where $z$ is the mean redshift of the group members. This correction factor has been checked using the G3C groups to produce a total luminosity consistent with the official \texttt{TotFluxProxy}.

The total stellar mass is another proxy for the total group mass. We take the \texttt{StellarMassesv19} DMU from \citet{2011MNRAS.418.1587T}, where stellar population synthesis is used to model the optical photometry of the GAMA galaxies. Because the modelling uses rest frame luminosities, which depends on distance, the stellar mass is expressed in units of $h^{-2}M_\odot$\footnote{Notice that this is only approximately true, because the stellar mass to light ratio, $M/L$, which is used to obtain the stellar mass, depends on age and is therefore specific to the choice of $h$. The stellar mass used here assumes $h=0.72$. }.Furthermore, for each group, we correct the total stellar mass by the same redshift dependent factor as the total luminosity. Notice that we do not apply the \texttt{fluxscale} correction here, which accounts for the missing flux from matched aperture photometry, because our results do not rely on the absolute stellar mass of the groups. This correction therefore does not affect our primary aim of splitting the groups into a few bins based on their ranking in mass.

\begin{figure}
\centering
	\includegraphics[width=0.45\textwidth]{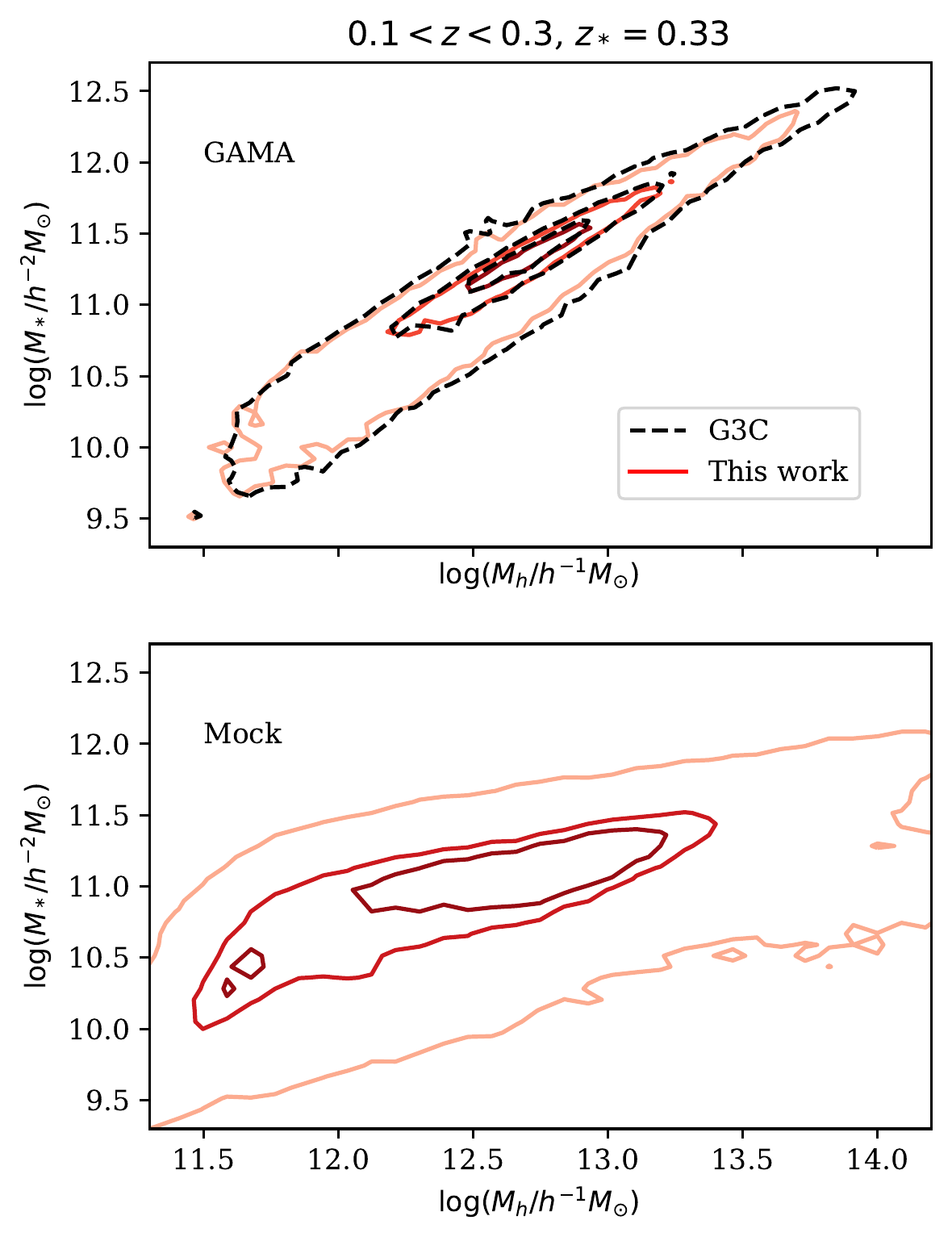}
    \caption{Upper panel: Correlation between the stellar mass, corrected to total by a factor $\exp(z^2/z_*^2)$ and the halo mass estimate derived from the total group luminosity mass proxy, together with a lensing-based absolute calibration from \protect\cite{jr:han}, for the GAMA groups with two or more members between redshifts $0.1<z<0.3$. The contours denote 95\%, 50\%, and 20\% of the total sample. The solid lines show the groups used in this work using the group finder algorithm in \protect\cite{jr:katarina1}, and the dashed lines show the the official G3C groups \protect\citep{jr:robotham}. Lower panel: The same relation for the mock catalogue. In this case, $M_h$ is not estimated from the luminosity, but directly taken as the arithmetic mean host halo mass of the group member. The difference in the distributions indicates that the stellar populations in the mock data are not entirely realistic, but it also warns us that the exact values of halo mass corresponding to the different GAMA group subsets must be treated empirically, and should not be treated as being known precisely.}
    \label{fig: stellar_lensing_mass}
\end{figure}

The calibration of the total stellar mass and the halo mass from weak lensing of the GAMA groups is shown in Fig.~\ref{fig: stellar_lensing_mass} for the official G3C groups from the \texttt{G3CFoFGroupv09} DMU (dashed line) and the group catalogue used in this work (solid line). The contours show 95\%, 50\%, and 20\% of the total sample, and are highly consistent between the two group catalogues. We choose to divide groups into three stellar mass bins based on percentiles: the Low Mass (LM) bin consists of the least massive 40\% groups, the Medium Mass (MM) bin corresponds to the middle 50\%, and the High Mass (HM) bin contains the most massive 10\%. The signal-to-noise of high mass haloes is expected to be high, despite the low number in the HM bin. 

\subsection{Mocks}
\label{sec:mocks}

We include mock catalogues for two reasons: (1) to validate the RSD models and assess the bias on the recovered growth rate, and (2) to quantify the impact of cosmic variance via the construction of covariance matrices. We used 25 realisations of a lightcone mock catalogue based on the GALFORM semi-analytical galaxy formation \citep{jr:gonzalez-perez}. The catalogue exploits the Millennium Simulation \citep{2009MNRAS.398.1150B} with the WMAP7 cosmology. These mocks can be obtained from the Durham hosted Virgo-Millennium Database1\footnote{\url{http://virgodb.dur.ac.uk:8080/MyMillennium/Help?page=databases/gama_v1/lc_multi_gonzalez2014a}} \citep{2006astro.ph..8019L}. For more details regarding the mock catalogue, see \cite{jr:farrow2015}.
By Eq.~\ref{eq: f}, the fiducial value of growth rate at the mean redshift of the mocks, $z=0.195$, is $f_{\rm fid}=0.593$. 
The lightcone is constructed using the methods in \citet{jr:merson}, where, given an observer, the galaxy is placed at the epoch where it first enters the past lightcone of the observer. The galaxy trajectories are interpolated between snapshots. Each mock covers the five GAMA fields with the SDSS $r$-band apparent magnitude ${\texttt{SDSS\_r\_obs\_app}<21}$, and $z<0.9$. 

\begin{figure}
	\centering
	\includegraphics[width=0.47\textwidth]{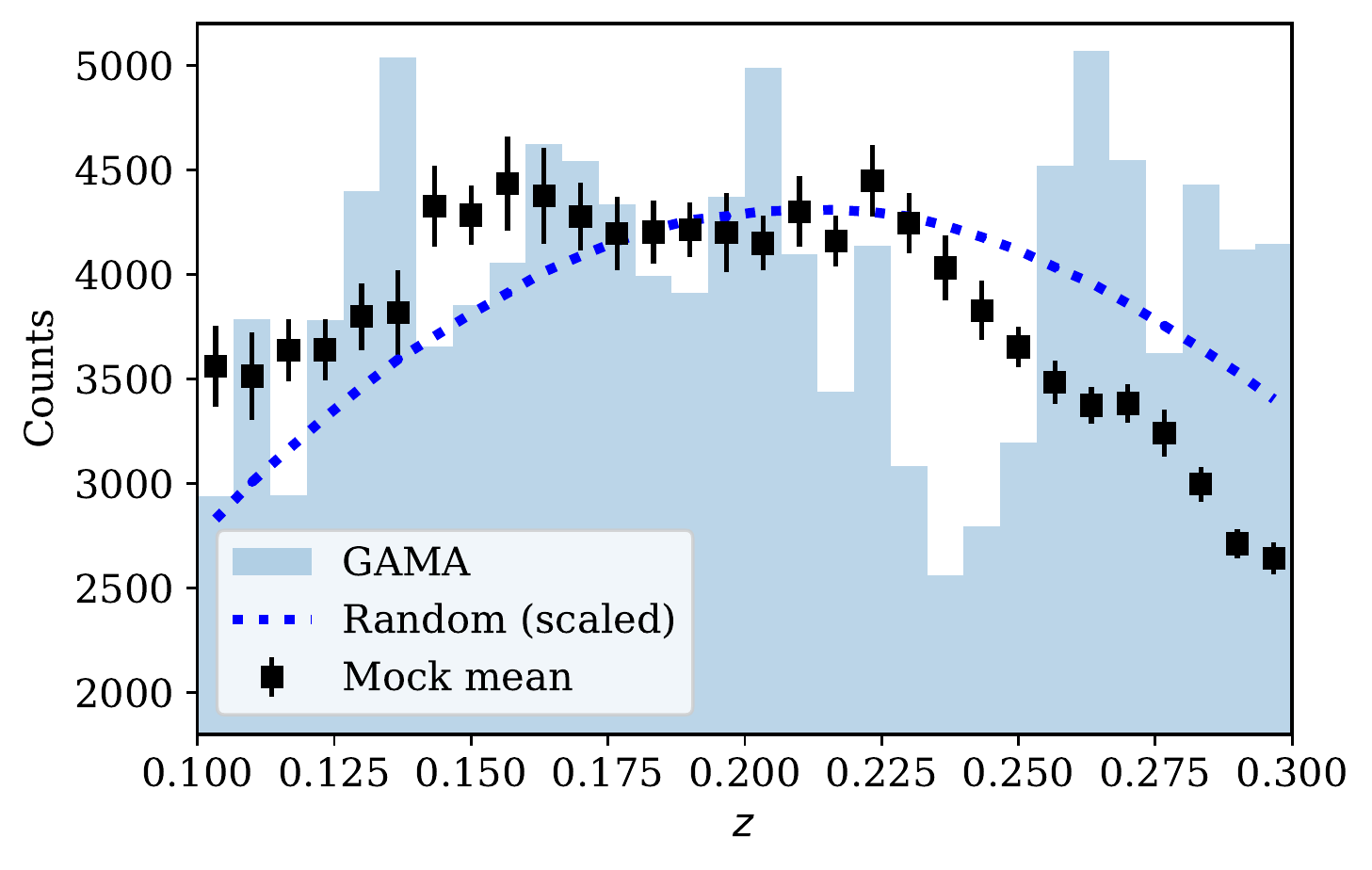}
    \caption{The mean redshift distribution of the 25 GAMA mocks (square) is offset from that of the random sample (dotted line). A random catalogue is created for the mocks to have matched redshift distribution as the mock mean. The redshift distribution of the GAMA galaxy sample is also shown (histogram) for comparison.}
    \label{fig: mock_gama_rand_zdist_0.1_0.3}
\end{figure}

We use galaxies in the G09, G12, and G15 fields and apply the same selection in redshifts $0.1<z<0.3$ and the apparent $r$-band magnitude cut ${\texttt{SDSS\_r\_obs\_app}}<19.8$. We also apply the same survey mask generated using the random catalogue. The masked areas are obtained by binning random galaxies in each field with an average of $\sim2000$ counts in each bin. Pixels with counts smaller than five times the Poisson noise are masked. The total masked area in the three fields is about $0.14\,{\rm deg}^2$. Because the mock redshift distribution is not matched exactly with GAMA data and random (see Fig.~\ref{fig: mock_gama_rand_zdist_0.1_0.3}), we create a random catalogue for these mocks by down-sampling the random catalogue for the GAMA data, such that the $n(z)$ matches the mean of 25 mocks.

The red and blue subsamples for the mean of the mocks are separated by the empirical line given by 
\begin{equation}
g-i=0.46+3.2z,
\end{equation}
as shown in the lower panel of Fig.~\ref{fig: color_split_gama_mock}. The line is chosen to go through the green valley of the mock galaxy $g-i$ colour. The GAMA galaxies have a more concentrated red sequence overlapping with an extended blue population, without a distinct green valley in between. On the contrary, the mocks have a broader red population which is well separated from the blue population by a green valley. Since the mock catalogues have more distinctive separation for the two populations, we find the corresponding colour cut in the GAMA data by matching red and blue fractions in the two catalogues for 20 redshift bins in $0.1<z<0.3$. The cut is smoothed by fitting a second order polynomial, as shown in the upper panel of Fig.~\ref{fig: color_split_gama_mock}. 

\begin{figure*}
	\includegraphics[width=0.7\textwidth]{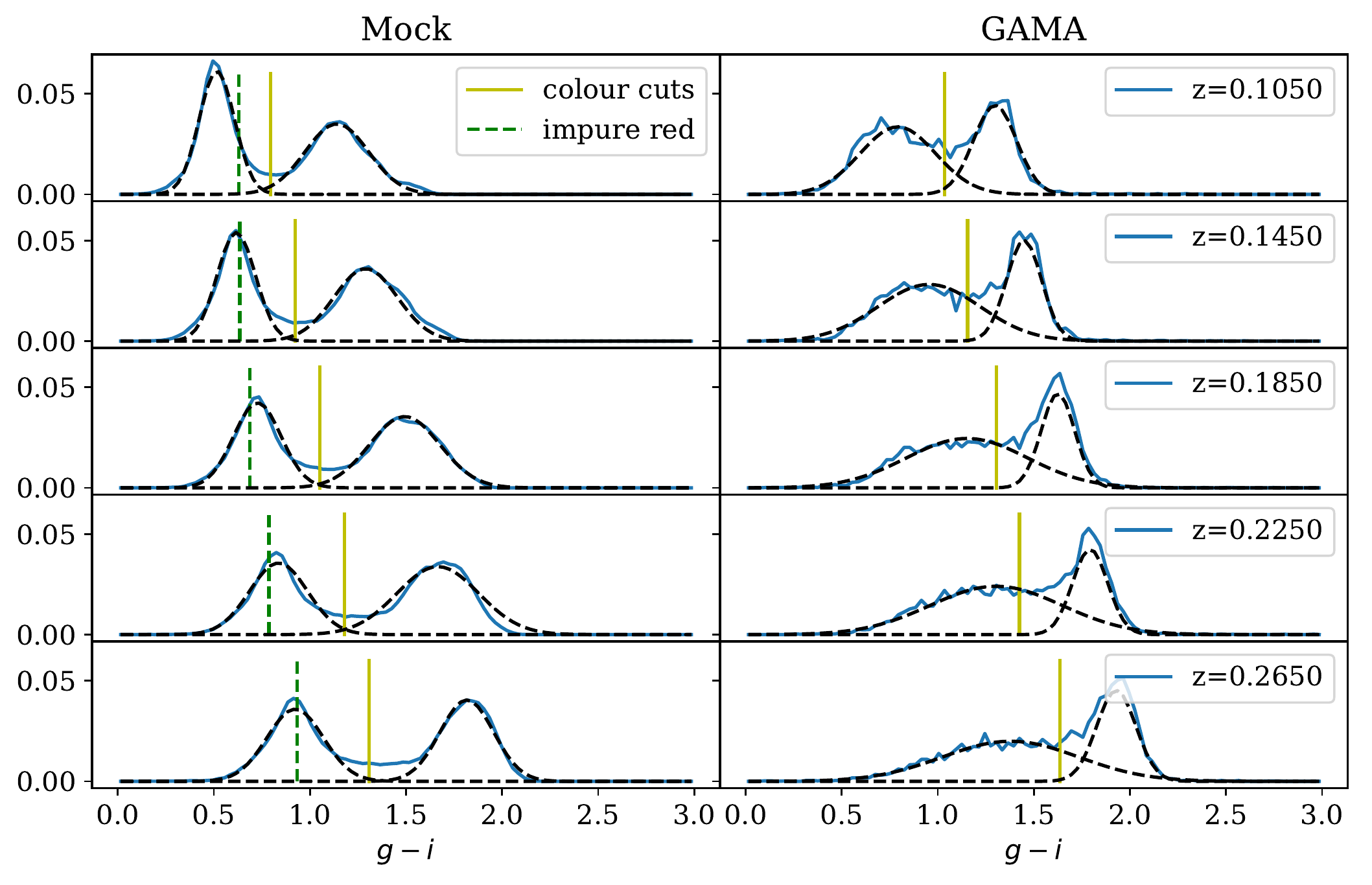}
    \caption{The $g-i$ colour distribution of GAMA and 25 mocks combined at 5 of the 20 redshift bins in $0.1<z<0.3$. The black dashed lines are double Gaussian fits to the distributions, characterizing the blue and red populations. The yellow cuts show a linear cut in the green valley in the mocks, and the corresponding cuts in GAMA which give the same red and blue fraction. The green dotted cuts in mock is the cut for the impure red sample.}
    \label{fig: colour_zbin_cuts}
\end{figure*}

The contamination of the red and blue sub-samples in the GAMA data resulting from the colour cut is quantified in the following way: for each redshift bin, the red and blue sub-samples are fitted by a double Gaussian. It is a reasonable fit except for the green valley in the mocks, as shown in Fig.~\ref{fig: colour_zbin_cuts}. Given a colour cut, the contamination of the red sub-sample is defined as the area under the blue Gaussian over the area under the red Gaussian, and similarly for the contamination of the blue sub-sample. Clearly, GAMA data contain a contaminated red sample and a pure blue sample. Therefore, we create a contaminated red sub-sample using the mock catalogues by placing the mock colour cut such that extra blue galaxies are included with the same level of contamination as GAMA data. The contaminated red cut in the mocks (see Fig.~\ref{fig: color_split_gama_mock}) is smoothed by fitting a quadratic polynomial of the form 
\begin{equation}
g-i=2.43z^2+1.55z+0.388.
\end{equation}

For mock groups, the stellar mass is computed by the sum of \texttt{diskstellarmass} and \texttt{bulgemass} of all group members, and corrected by the same redshift-dependent factor as the data. We do not estimate the group halo mass from the same mass-luminosity relation in Eq.~\ref{eq: mh vs L}. Instead, we use the host halo mass of the mock galaxy directly. 
Because some haloes contain more than one galaxy, for each group, we test the largest, the arithmetic mean, and the median halo mass of the group member, and find that they give similar results. We also test using the sum of unique host haloes in the group. This increases the total group halo mass in the lower mass end, but does not affect the higher mass end. 
The stellar-halo mass relation of the groups using the total stellar mass and the arithmetic mean host halo mass of the group members is shown in the lower panel of Fig.~\ref{fig: stellar_lensing_mass}. It is clear that the mocks show a much larger scatter in the $M_h-M_*$ plane and the slope is smaller compared to data, i.e., at fixed stellar mass, the halo mass is larger. The total stellar mass of the mock groups is also smaller by about 0.5 dex compared to data. The clear difference between data and the mocks shows that estimating the halo mass from luminosity using Eq.~\ref{eq: mh vs L} is not very reliable. The luminosity is itself strongly correlated with stellar mass via the luminosity-mass relation, thus the upper panel of Fig.~\ref{fig: stellar_lensing_mass} does not show the true scatter of $M_h$ at fixed $M_*$ faithfully (or vice versa).
A comparison between the group and the halo catalogue in the GAMA mocks reveals that the 2-member groups have low fidelity, also discussed in \citet{jr:robotham}. This again emphasises the importance of using a consistent group finder algorithm between the GAMA data and the mock catalogues. The mock group catalogues are separated into three stellar mass bins based on the 40\%, 50\%, and 10\% percentiles as measured in the data.

\section{Cross-correlation measurements}

\subsection{Correlation  statistics}
\label{sec:measurement-statistics}

We estimate the 2-point correlation function by counting pairs of galaxies and randoms using the Davis--Peebles estimator \citep{jr:davispeebles}:
\begin{equation}
\hat{\xi}(r_p,\pi)=\frac{D_1D_2}{D_1R_2}-1,
\label{eq: D-P estimator}
\end{equation}
where the subscript $i=1,2$ denotes the two samples to be correlated (in case of auto-correlation, the same sample), $D_i$ denotes data, and $R_i$ denotes the corresponding random points. Each term in the equation (e.g., $D_1D_2$) is the normalised pair count between data and (or) random points, measured in bins of the pair separation $(r_p, \pi)$ defined below. 
For two objects located at $\mathbf{s}_1$ and $\mathbf{s}_2$, their separation is given by $\mathbf{s}=\mathbf{s}_1-\mathbf{s}_2$. The line of sight is defined along the mean position of the pair, $\mathbf{r}=(\mathbf{s}_1+\mathbf{s}_2)/2$. One can then decompose the separation into components parallel and perpendicular to the line of sight:
\begin{equation}
\pi=\frac{\mathbf{s}\cdot\mathbf{r}}{|\mathbf{r}|}; \quad r_p=\sqrt{\mathbf{s}^2-\pi^2}.
\end{equation}

The random catalogue $R$ captures various properties of the actual data, such as the survey mask and the sample redshift distribution $n(z)$, but has no spatial correlation. Thus, these estimators essentially measure the excess clustering of the data points compared to a random distribution. 
For the red and blue galaxy samples, the random $n(z)$ is modulated by the redshift-dependent red and blue fraction respectively (see Section~\ref{sec:GAMA galaxies and groups}). 

For the case of auto-correlations, the Landy--Szalay estimator \citep{jr:landyszalay} is known to be superior to the Davis--Peebles approach, and there is a natural generalisation to cross-correlation:
\begin{equation}
\hat{\xi}(r_p,\pi)=\frac{D_1D_2-D_1R_2-D_2R_1+R_1R_2}{R_1R_2}.
\label{eq: L-S estimator}
\end{equation}
However, implementing this estimator would require a random catalogue for galaxy groups in different mass ranges, and we prefer to avoid this complication. 
In contrast, the Davis--Peebles estimator requires a random catalogue for only one of the populations being correlated.
Both \cite{jr:mohammad} and \cite{2021MNRAS.506...21R} estimated cross-correlations using a form of Landy--Szalay where 
$R_2$ was replaced by $R_1$, but this has no justification, and will yield incorrect results when the selection functions of the two tracers are very different.
We measured the galaxy auto-correlations using both estimators, and found negligible difference for our sample. Throughout the analysis, the size of random catalogue used is 20 times that of data. 

The 2D correlation functions are measured out to a maximum scale of $40\mpcoh$ for both the $r_p$ and $\pi$ directions in bins of $1\mpcoh$. 
This maximum scale is chosen due to the limited volume of the GAMA survey. It may appear to be a concern that at this scale perturbation theory starts to break down (thus the scale is often chosen as the cut-off scales for larger data sets). However, we shall show later that empirical models based on linear theory can still give relatively unbiased results below
this scale. For the
halo streaming model, which we will elaborate in Section~\ref{sec:model2}, we use a 1-halo template to absorb the deviation of non-linear clustering from the perturbative 2-halo term.
Although in principle, $r_p$ and $\pi$ should be measured in the range $[-40,40]\mpcoh$, in practice, pair counts in the positive and negative bins are combined and our correlation functions have two mirror planes of symmetry.
This is the standard practice for $r_p$, because the correlation function is symmetric around the transverse direction. 
Along the line of sight, positive and negative $\pi$ measurements can in fact be distinctive in cross-correlations due to secondary gravitational effects. For example, gravitational redshift can give rise to a non-vanishing dipole between two samples that differ significantly in mass \citep{2014PhRvD..89h3535B,2011Natur.477..567W, Cai2017, Beutler_2020}. However, given the size of the sample, we shall not investigate this issue in the present study. The combination of the $\pi$ bins also improves the signal-to-noise for our measurements. 

\begin{figure*}
\centering
\includegraphics[width=\textwidth]{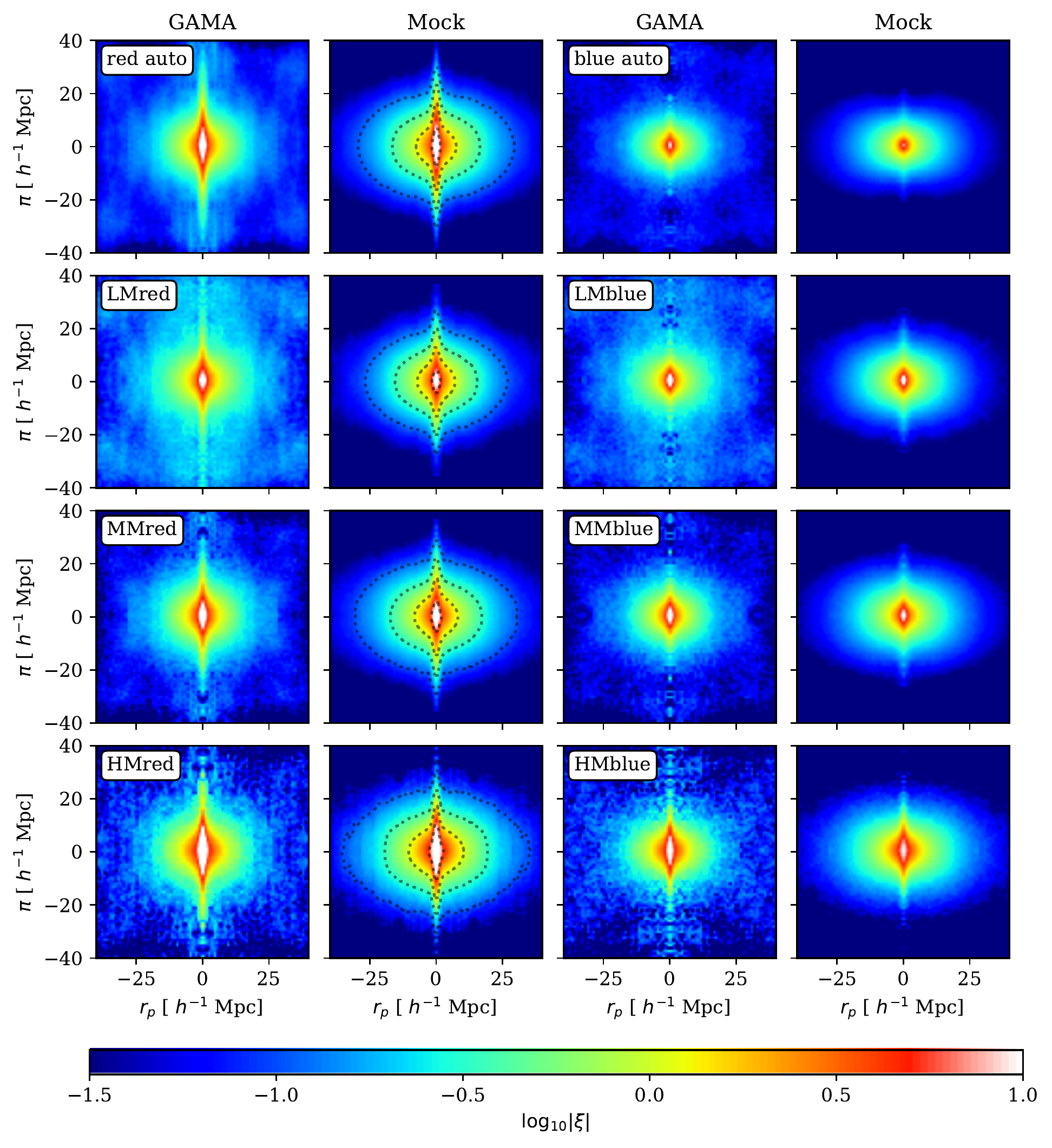}
\caption{False colour images of auto- and cross-correlation functions in redshift space for the actual GAMA data and the corresponding average over the set of 25 GAMA mocks. $r_p$ denotes transverse separation; $\pi$ is radial separation. LM, MM, and HM denote the three group mass bins. A number of trends are apparent: both the bias (amplitude of clustering) and the small-scale Finger of God (FoG) dispersion increase with group mass, and are larger for red galaxies than for blue. The mock contaminated red sample is shown as dotted contours on the second column, with $\log_{10}|\xi|=\{-1.0, -0.5, 0, 0.7\}$.}
\label{fig:all2d_mocks}
\end{figure*}

\subsection{Results}
\label{sec:measurements}

In the following analysis, we will refer to the group subsamples as LM (low mass bin), MM (medium mass bin), and HM (high mass bin). We thus have six configurations for the group-galaxy cross-correlations: LMred, MMred, HMred, LMblue, MMblue, and HMblue.
In addition, we also measure the red and blue galaxy auto-correlations. The inclusion of galaxy auto-correlations in the analysis helps in breaking the near degeneracy between $b_{\rm gal}$ and $b_{\rm grp}$. 
Ideally, one would also include the auto-correlations for the group catalogue. These are excluded here: as mentioned above, we do not construct random versions of the group catalogue.

Fig.~\ref{fig:all2d_mocks} shows the red and blue galaxy auto-correlations (top row) and the cross-correlation functions for the six configurations (lower three rows) measured from the GAMA data (first and third column), and the corresponding mocks average (second and forth column). 
The first noticeable feature is that the red configurations (left two columns) have larger clustering signals compared to the blue ones (right two columns). This is most obvious in the two galaxy auto-correlations. On relatively large scales, the squashing is also stronger in the blue configurations. This effect is controlled by the Kaiser distortion parameter $\beta=f/b$ (see Section~\ref{sec:model1} below), and is expected to be stronger for samples with a smaller bias $b$ (vice versa), given that the growth rate $f$ is fixed. The fact that red galaxies have a larger galaxy bias compared to blue galaxies implies that red galaxies are preferentially associated with more massive dark matter haloes, in agreement with other studies \citep[e.g.][]{jr:guo2014, 2016MNRAS.457.3200M, 2021A&A...653A..82B}. 
On smaller scales, the red configurations also show a much more prominent FoG signal compared to the blue ones. This is intuitively sensible because more massive haloes are associated with a larger velocity dispersion. 
Comparing the correlation functions across different groups for a specific galaxy selection, we see a similar trend on both small and large scales: with the increase in group mass, a larger clustering amplitude, group bias, and FoG effect are observed. Notice the high signal-to-noise at small scales in the HM groups, despite that the sample size in this mass range is only about $1/4$ of the other mass ranges. These observations confirm that the identification of the galaxy groups, as well as the separation of group masses based on the effective halo mass (total luminosity) are successful for the purpose of this study.

The agreement between the mock average and the GAMA data is good in general -- the same trends in galaxy colour and group mass are captured. In regions where $r_p$ is close to zero, the mock average seem to produce weaker clustering compared to the actual data in the blue configurations. The match in the red configuration, on the other hand, is excellent. 
The mock contaminated red sample is also shown as dotted contours (second column). The inclusion of extra blue galaxies has the effect of slightly reducing the overall amplitude in this contaminated sample compared to the pure red sample.
On larger scales, the signal in data is noise and cosmic variance dominated. This is most noticeable in the LM subsamples, where the signal greatly exceeds the mock average on $r\geq20\mpcoh$. Inspecting the measurement in each mock sample, this level of fluctuation in the data is expected.
It should be noted that cosmology adopted in the mock catalogue is $\Omega_m=0.27$, which is lower than the current constraint from {\em Planck},  $\Omega_m=0.315\pm0.007$ \citep[][]{2018:planckpars}. Thus, one may expect some difference in clustering between the mock average and the data. However, given the noise in the GAMA data, a percent-level shift in the growth rate $f\propto\Omega_m^{0.55}$ is hard to discern. 
It is also found in \cite{jr:farrow2015} (e.g. their Figure 8) that the mock can capture similar clustering trends as the GAMA data, when split into bins of redshift and stellar mass. Notice that there are significant deviations at small scales ($r_p<1\mpcoh$) in the shape of the projected correlation functions, but these scales are not explored in this analysis.

\section{RSD models}
\label{sec:models}

\subsection{Quasilinear dispersion model}
\label{sec:model1}

To describe the RSD in galaxy density field, one can extend Eq.~\ref{eq: kaiser} by including the galaxy bias $b$:
\eqb
P^s_g(k,\mu)= b^2P^r(k)\,(1+\beta\mu^2)^2, 
\label{eq:galaxy_ps}
\eqe
where $\beta =f/b$ is often referred to as the distortion parameter. 
The above formalism is valid for galaxy auto-correlation, but it is straightforward to generalise to cross-correlation:
\eqb 
P_c^s(k,\mu)=\frac{b_{\rm gal}^2}{b_{12}}(1+\beta_{\rm gal}\mu^2)(1+b_{12}\beta_{\rm gal}\mu^2)P^r(k),
\label{eq: model1 P}
\eqe 
where $b_{\rm gal}$ is the galaxy bias, and $b_{12}$ is the ratio between galaxy and group bias:
\eqb
b_{12}\equiv b_{\rm gal}/b_{\rm grp}.
\eqe

The 2-point correlation function is the Fourier transform of the power spectrum. It is convenient to express the correlation functions in terms of Legendre polynomials $\mathcal{P}_\ell(\mu)$ with $\ell=0,2,4$ \citep{jr:hamilton}:
\eqb 
\xi^s_g=\xi_0(r)\mathcal{P}_0(\mu)+\xi_2(r)\mathcal{P}_2(\mu)+\xi_4(r)\mathcal{P}_4(\mu).
\label{eq: multipole expansion}
\eqe 
In this expression, the coefficients of the Legendre polynomials, $\xi_{0}(r)$, $\xi_{2}(r)$, and $\xi_{4}(r)$ are referred to as the monopole, quadrupole, and hexadecapole. In linear theory, only even modes are present up to the forth order because of the RSD effect modifies the power spectrum by the factor $(1+\beta \mu^2)^2$. The specific form of these multipoles are computed by \citet{jr:hamilton} for auto-correlation, and \citet{jr:mohammad} for cross-correlation. We summarise these formulae in Appendix~\ref{appendix: model linear}.

The FoG effect is accounted for by a convolution of the correlation function with some distribution of the non-linear random peculiar velocity along the line of sight \citep{1994MNRAS.267.1020P}.  $N$-body simulations show that the actual distribution is non-Gaussian \citep[e.g.][]{1996MNRAS.279.1310S,Scoccimarro2004, Cuesta-Lazaro2020}. Thus, we adopt:
\begin{equation}
D(\pi)=\frac{1}{\sqrt{2}\sigma_{12}}\exp\left(-\sqrt{2}H_0 \pi/\sigma_{12}\right),
\label{eq: vdispersion}
\end{equation}
where $\sigma_{12}$ is the pairwise velocity dispersion. In Fourier space, this function takes the form of a Lorentzian function, $\tilde{D}(k\mu)=[1+(k\mu\sigma_{12})^2/2]^{-1}$, which damps the high-$k$ modes of the anisotropic power spectrum.

At quasilinear scales, non-linearity may introduce systematic biases in the inferred cosmological parameters \citep{jr:delatorre2012}. There are multiple challenges in extending the model beyond linear regime. There, the peculiar velocities can be large, and the formalism described breaks down at linear order. Nonlinearities alter the small scale shape of the matter power spectrum and correlate the density and velocity fluctuations. Accounting for these effects requires higher order expansion in the Perturbation Theory and the inclusion of the velocity spectrum, $P_{\theta\theta}(k)$, and the density-velocity cross spectrum, $P_{\delta\theta}(k)$, e.g. the TNS model by \citet{jr:tns}. Galaxy bias can also be nonlinear and stochastic on small scales \citep{jr:dekellahav}. Furthermore, the approximate velocity dispersion in equation~\ref{eq: vdispersion} fails to fit auto-correlation data on the smallest scales. More elaborate velocity distributions are proposed by e.g., \citet{jr:reidwhite}; \citet{jr:zuweinberg}; \citet{jr:bianchi} based on simulations.

One simple approach is the replacement of the linear power spectrum in the linear Kaiser model (Eq.~\ref{eq:galaxy_ps}) by the non-linear power spectrum. This is reasonable because the redshift space power spectrum should match that in real space at $\mu=0$. \cite{jr:blake2011} showed that this combination is actually among the best-performing RSD models when fitting down to $k_{\rm max}=0.2\hompc$ with fixed cosmology\footnote{Although it should be noted that if the model could introduce bias to $\Omega_m$ if the cosmology is not fixed, as shown in \citet{2012PhRvD..86j3518P}.}.
For this model, we adopt the non-linear power spectrum from {\sc Halofit} \citep{jr:smith2003,Halofit2}. In the nonlinear regime we should in principle allow for a scale-dependent bias. But in practice it is a good approximation to assume that the nonlinear galaxy and matter power spectra are in a constant ratio (see \citejap{2019MNRAS.487.3870C}).
In the following analysis, we refer to this model as the `quasilinear dispersion' (QD) model.

\subsection{RSD in the halo model}
\label{sec:model2}

The main deficiency of the QD model is that it does not address post-linear couplings between density and velocity, which will modify the simple Kaiser angular anisotropy. There is an extensive literature of attempts to improve such modelling, based on various forms of perturbation theory. The model of \cite{jr:tns} is widely used, although more recent efforts have concentrated on the Effective Field Theory approach. This adds additional terms dictated by symmetry in a way that can also capture bias effects, including non-linearity and non-locality (e.g.  \citejap{2012JHEP...09..082C}; \citejap{2015JCAP...11..007S}; \citejap{2020JCAP...05..005D}). These results are impressive, but have the limitation that they are presented in Fourier space and are not reliable beyond $k\simeq0.3\hompc$. For a robust prediction of correlation functions, we need a formalism that still behaves correctly in the large-$k$ limit.

For this reason, we have developed a model that seeks to access the highly nonlinear regime by using the halo model. In real space, this involves correlations that count pairs of galaxies in the same halo or in different haloes:
\begin{equation}
\xi(r) = \xi_{1h}(r) + \xi_{2h}(r).
\end{equation}
The 1-halo term is determined by the form of the halo density profile, and the 2-halo term is close to a linearly biased version of the matter two-point function. The bias in turn is determined by the halo occupation number, $N(M)$, of galaxies in haloes as a function of their mass. This halo model has proved a highly effective way to understand the relation between the clustering of galaxies and of mass \citep{HM_Seljak, HM_PS, HM_CS}, and for the case of dark matter alone has led to the highly precise HALOFIT framework \citep{jr:smith2003,Halofit2}.

The halo-model separation into two independent pair contributions must also apply for the redshift-space correlations, namely
\begin{equation}
\xi(r_p,\pi) = \xi_{1h}(r_p,\pi) + \xi_{2h}(r_p,\pi),
\end{equation}
but it should be clear from the outset that the 1-halo and 2-halo contributions would be expected to have rather different anisotropy signals. The characteristic quadrupole plus hexadecapole Kaiser distortion arises from the coherent component of the velocity field, and this will apply to the 2-halo term only, since pairs from within the same halo are unaffected by bulk motion of the halo. This redshift-space decomposition using the halo model was advocated by \cite{2017JCAP...10..009H}, who invested much effort in trying to predict the two distinct components using perturbation theory. Our work bears some resemblance to their approach, with two distinct differences: we work directly in configuration space, and we base the 1-halo term on empirical simulation results, rather than attempting to calculate it a priori.

A particular point to clarify in this decomposition is the treatment of Fingers of God. 
Random motions within a halo are treated in the dispersion model by a radial convolution -- but in fact the appropriate convolution will be different for the 1-halo and 2-halo terms. 
The main reason for this is that the 1-halo and 2-halo terms weight contributions as a function of halo mass differently, with a higher weight given to high-mass haloes in the 1-halo term (see e.g. equations 8 \& 10 of \citejap{HM_Seljak}). Since the pairwise dispersion $\sigma_{12}$ increases with halo mass, we expect larger FoG effects to apply to the 1-halo term. This is further complicated by the existence of central and satellite galaxies, since the weighting of these is different in the 1-halo and 2-halo terms. For example, suppose each halo contains either just a single central or one central and one satellite, where the velocity dispersion of satellites is $\sigma$. The 1-halo contribution must pair a central with a satellite, so the pairwise dispersion is $\sigma$. But the 2-halo term can also pair centrals with centrals (assumed to have negligible pairwise dispersion -- although \citejap{2014MNRAS.444..476R} showed that the actual pairwise velocity could be up to $30\%$ of $\sigma$) and satellites with satellites (pairwise dispersion $\smash{\sqrt{2}\sigma)}$, so the average rms pairwise dispersion depends on the fraction of haloes that contain a satellite. If most haloes are central-only (as in BOSS CMASS, for example), the pairwise dispersion for the 2-halo term will be $\ll\sigma$. In the opposite direction, one can argue that the velocity field of haloes will contain some stochastic component in addition to the coherent velocities that generate the Kaiser distortion.

With this perspective, an improved simple model for the cross-power between tracers $a$ and $b$ would be as follows:
\begin{equation}
\eqalign{P_{ab}(k,\mu)
= &P_{1h}(k) \, D_1(k\mu) \cr
&+ b_a b_b P_{\rm lin}(k) \,(1+\beta_a\mu^2)(1+\beta_b\mu^2) \,D_2(k\mu).
}\label{eq:2damp}
\end{equation}
Leaving aside the 1-halo term for the moment, one way in which we can seek to improve this expression further is in terms of quasilinear effects on the 2-halo term. A first requirement is that the real-space spectrum (at $\mu=0$) should have the full nonlinear form.
When discussing the dispersion model, we achieved this by replacing $P_{\rm lin}$ by the nonlinear spectrum. In the halo model, we should not do this, since the 2-halo term in real space is close to linear theory, and the 1-halo term supplies most of the nonlinear corrections \citep{jr:smith2003}. We do however adopt the HALOFIT 2-halo term, with scale-independent bias, as the best model for the real-space 2-halo term.

The next step is to seek improvement in the density-velocity coupling that leads to the Kaiser distortion factors. An attractive approach here is the streaming model (e.g. \citejap{Fisher1995}; \citejap{2016JCAP...12..007V}), in which we consider the quasilinear relative velocity distribution as a function of pair separation, and use this to transform to redshift space while exactly conserving pair counts. The details of the construction of this model are given in Appendix \ref{appendix: model}. As with the linear model for the 2-halo term, there are three main free parameters, the tracer biases and the growth rate: $(b_a,b_b,f)$. This assumes that the mass power spectrum is known exactly, whereas it depends on all fundamental $\Lambda$CDM parameters. The main variation of the power is $\propto\sigma_8^2$, so it is common to factor out this degree of freedom and take the main RSD parameters to be $(b_a\sigma_8,b_b\sigma_8,f\sigma_8)$. However, there is a weaker further dependence on $\sigma_8$ when we adopt the HALOFIT prediction of the 2-halo matter power spectrum, rather than taking this to be pure linear theory (although the difference is not important in practice). 

In addition to these three main RSD parameters, we have parameters connected to the FoG damping. As described earlier, it is conventional to model FoG effects by radial convolution, taking the velocity PDF to be a Lorentzian and using a velocity dispersion as the single free parameter. But in the present context, it is important to be clear that the empirical evidence for the Lorentzian form comes mainly from the 1-halo term. This is $D_1(k_z)$ in Eq. \ref{eq:2damp}; but we want the effect on the 2-halo term, $D_2(k_z)$. We have argued that this will be characterised by a different dispersion, but in addition there is no strong reason to assume it will have a Lorentzian form. As a more general alternative, we considered a modified Lorentzian:
\begin{equation}
    D_2(k\mu) = \left(1+ (k\mu\sigma_{12})^2/2\gamma\right)^{-\gamma}
\end{equation}
and experimented with different values of $\gamma$. But in practice the results were rather insensitive to the choice of this parameter, so we retained the Lorentzian $\gamma=1$. 
It is shown in e.g. \cite{Scoccimarro2004, jr:bianchi, Cuesta-Lazaro2020} that the shape of the PDF is only relevant where the correlation function changes significantly over a scale comparable to the width of the smoothing function.
But the issue of the exact form of the PDF for FoG corrections to the 2-halo term is a problem that merits further study.

In summary, we therefore have two models with similar real-space correlations, but different degrees of RSD: (1) QD: quasilinear dispersion model and (2) HS: halo+streaming model.
Both of these converge to the linear Kaiser model on large scales, so what is of interest is the smallest scale to which their predictions are reliable. 
We will assess these by comparison with mock data.

\begin{figure*}
\centering
	\includegraphics[height=0.3\textwidth]{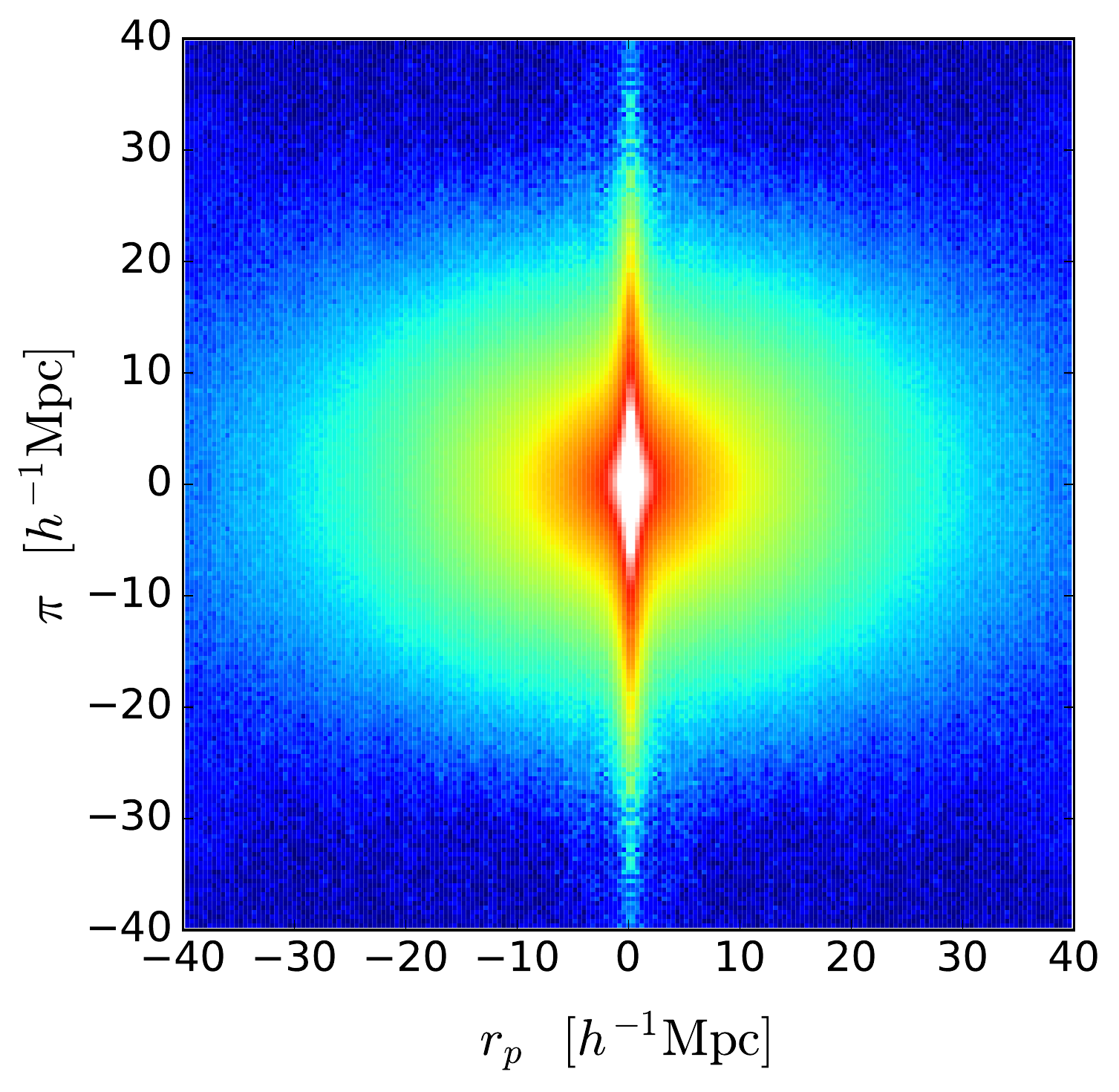}
	\includegraphics[height=0.3\textwidth]{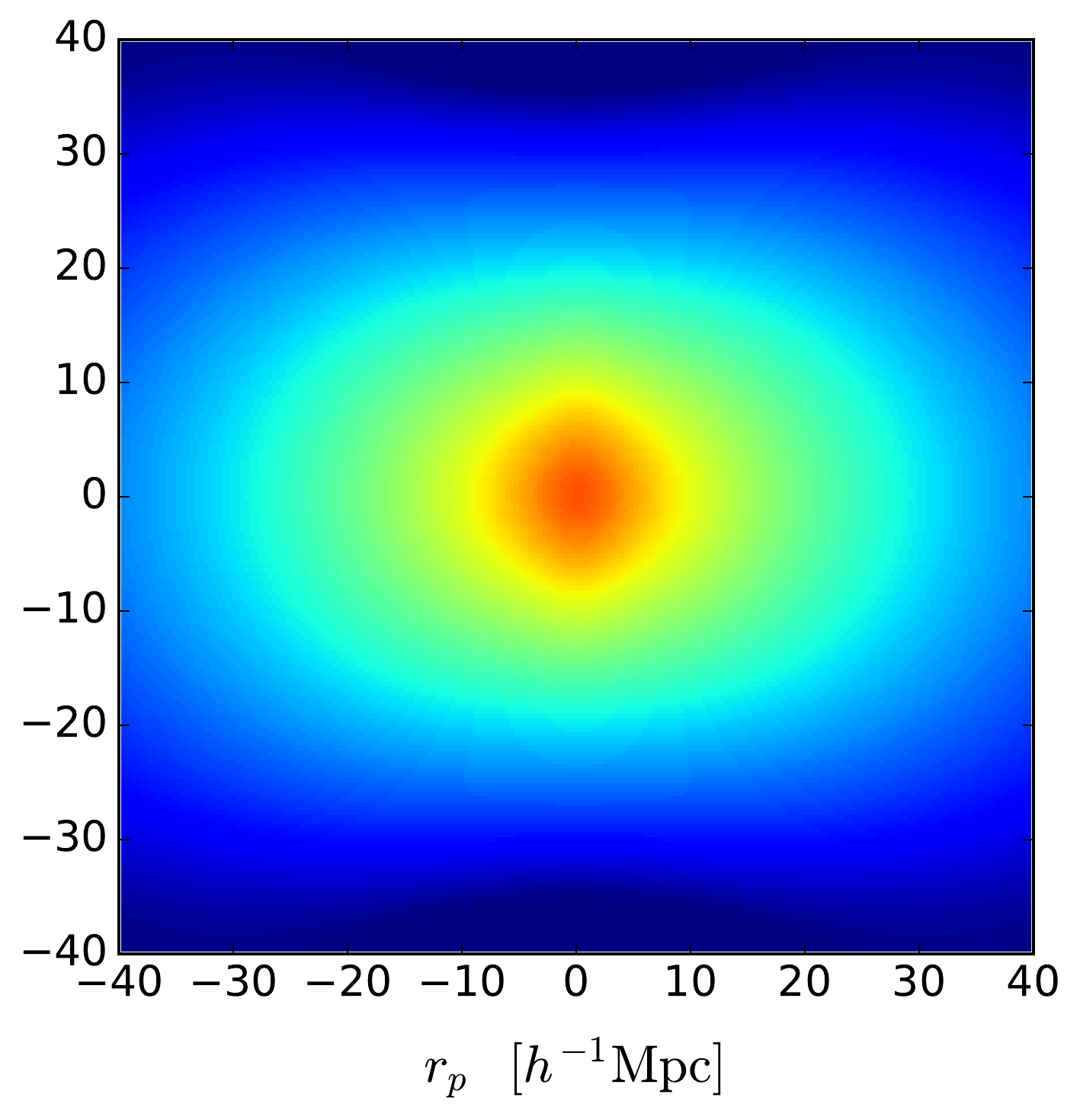}
	\includegraphics[height=0.3\textwidth]{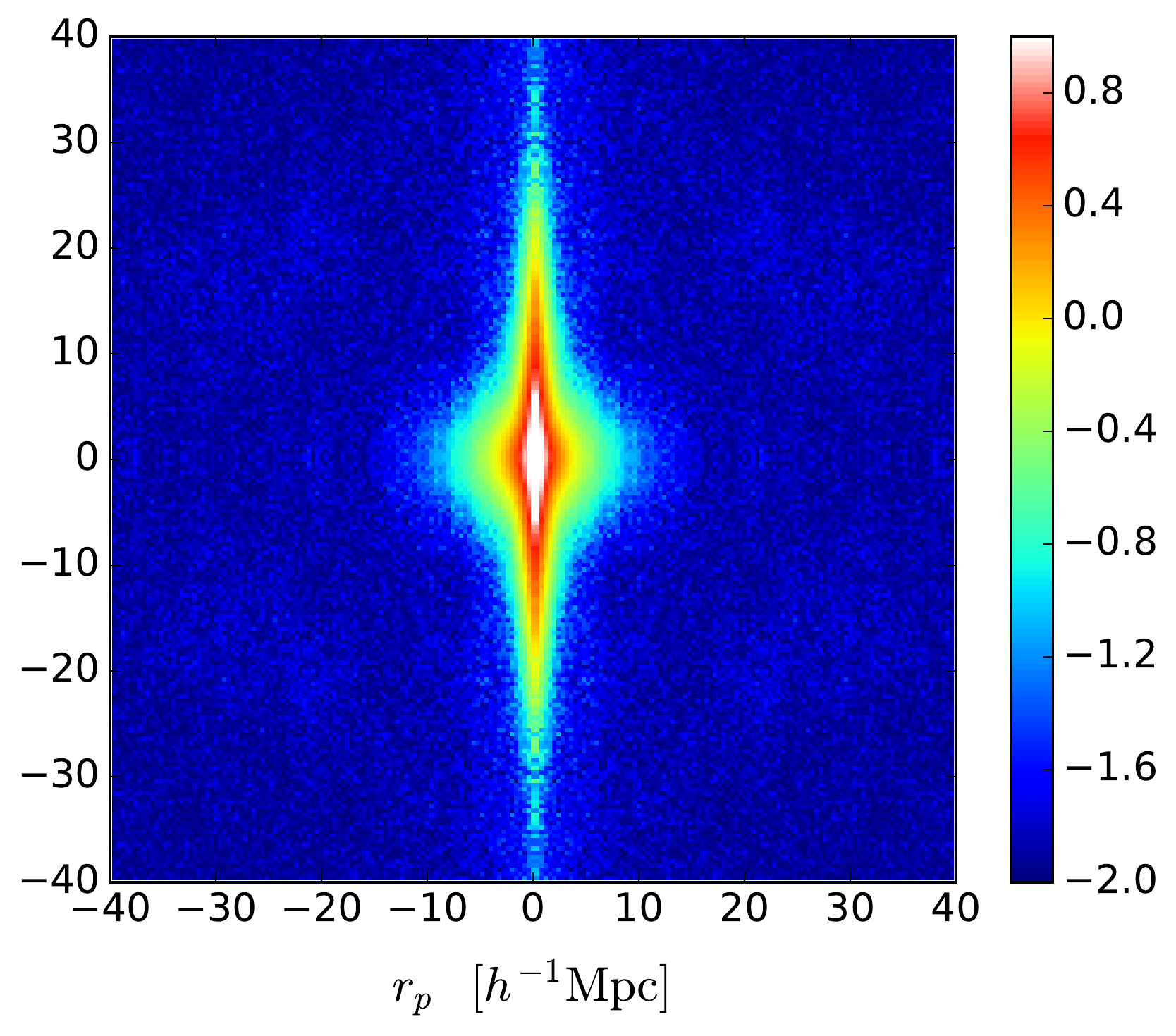}
    \caption{Illustrating the decomposition of the measured mock correlation data (left panel) into a 2-halo fit (middle panel) and an empirical 1-halo term in the form of the residual of the fit (right panel), for the particular case of red-MM cross-correlation. The 2-halo term is computed using the streaming model, and is matched to the data at radii $r>10\mpcoh$, with the additional criterion that $r_p>3\mpcoh$.}
\label{fig:1hplot}
\end{figure*}

\subsubsection{The 1-halo term}

The real-space 1-halo term can in principle be computed in the usual halo model framework, given the occupation numbers for the tracers and the halo radial profile. 
But there is also a case for taking an empirical approach, given that the real-space correlations are in principle observable directly, in a manner free of RSD effects, via the projected correlation function $w_p(r_p)$. One might for example model the real-space 1-halo term by a
power-law of free amplitude and slope, or via an NFW profile.

But whatever approach is taken in real space, there is then the question of how the 1-halo term appears in redshift space. As described above, the simplest approach is to assume that the transition to redshift space consists of a radial convolution with a single FoG function. However, it is not hard to see that this must be an oversimplification. The 1-halo term arises from random orbital velocities within the halo, but the velocity dispersion is unlikely to be constant. If for example we consider the case of isotropic orbits, then the dispersion would need to fall to zero at the virial radius of the halo, beyond which the density is assumed to vanish. 

Here we address this concern directly by using the mocks. Given a hypothesis for the 2-halo term, we can subtract the 2-halo prediction from the mock data to obtain an empirical $\xi_{1h}(r_p,\pi)$ that sums with the 2-halo term to give exactly the mock data (specifically, we apply this approach to the average of all the mocks). The 2-halo term can be deduced by fitting to the mock data in a regime where we assume the 1-halo contribution to be negligible. The exact cuts adopted in the process are not critical; in practice, we chose to match to the
data at radii $r>10\mpcoh$, with the additional criterion that $r_p>3\mpcoh$. The operation of this procedure is illustrated in Figure
\ref{fig:1hplot}. The resulting residual 1-halo term is clearly well localised near the origin, and indeed it can be seen that the RSD effects in the 1-halo term are complicated, with the FoG effect being largest at $r_p=0$, whereas the function appears more isotropic close to its outer limit
at $r_p\simeq 5\mpcoh$. This interesting behaviour is clearly worthy of being modelled in detail, but we shall not do that here.

We now have a decomposition of the redshift-space correlations that by construction exactly matches the average of the mocks. However, each mock realization will be different, as will be the real data, so can these different datasets be fitted in this framework? The 2-halo term is already parameterised, and these parameters can be varied for any given dataset. But the 1-halo term must also have some variation. Our approach is to assume that the mocks are sufficiently realistic that the effective 1-halo term in any given case will be close to the mock average, and that the difference can be captured by two nuisance parameters:
\begin{equation}
   \xi_{1h}( r_p, \pi )  \to  \alpha \xi_{1h}( r_p, \eta \, \pi ). 
\end{equation}
In other words, assume that we have roughly the right functional form, but that the amplitude may be off (scale by $\alpha$), and that the FoG strength may be off (stretch in the radial direction by $1/\eta$). 
Physically, the amplitude parameter $\alpha$ can be related to the way in which galaxies populate haloes: there may be different numbers of satellite galaxies in a given halo compared to the mock, leading to a different small-scale clustering signal. We have previously seen that an empirical rescaling of the 1-halo amplitude can yield an accurate fit to correlation data in real space (\citejap{2021MNRAS.501.1481H}).  The $\eta$ parameter attempts to capture the velocity dispersion of the galaxies in the halo: a smaller $\eta$ produces a larger FoG effect. Again, this can be understood in terms of an uncertain halo occupation, which can alter the mean mass of the haloes that contribute the 1-halo term.

As we show below, this approach is able to succeed in matching the individual mock realizations, and so we see no reason not to apply the same model to the real data. We emphasise that we do not need to assume that the mocks are completely realistic, as long as they are qualitatively similar to reality. The reliability of this approach can be judged by whether or not the fitted values of $\alpha$ and $\eta$ are close to unity (as indeed turns out to be the case). 
We only consider this minimal set of two empirical nuisance parameters in the current analysis; for forthcoming large data sets of higher statistical precision, more parameters may be required in order to make the model acceptably accurate. Eventually, we will need to validate the model by deriving 1-halo templates from a given set of mocks and showing that they can fit data derived from mocks produced according to different assumptions. We intend to pursue this high-precision robustness test in a future study.

\subsection{Fitting methodology}

\subsubsection{Covariance matrix and likelihood inference}

In the above discussion, we have not been explicit about exactly what it means to fit the averaged mock data. In principle, one would like to have an understanding of the errors on the data, so that the likelihood can be computed as a figure of merit that is used to optimise the fit. For an individual dataset, this can be done in the standard way by using an ensemble of mocks to estimate the covariance matrix of the data, and then appealing the to central limit theorem to compute the likelihood in the Gaussian approximation. For fitting the stacked mocks, the appropriate covariance matrix is less obvious, but in any case it is less important to have a likelihood in that case, where the aim is simply to estimate a 1-halo contribution as a basis for further modelling. We are not interested in placing errors on the best-fitting parameters of the 2-halo term, for which a likelihood would be required. In practice, therefore, we took the simple approach of seeking a least-squares fit in $\ln(1+\xi)$ to the mock average. The exact figure of merit chosen is unimportant as regards the 1-halo residual.

The covariance matrix for a single data realization is most often estimated in one of two ways: either directly via the scatter over a number of mock realizations, or via Jackknife resampling of a single realization. Both of these approaches have their limitations, but the best strategy is when they are combined: an expanded set of mock realizations is created by Jackknife resampling of each one, yielding an improved estimate of the covariance matrix \citep{2021MNRAS.503...59A}.
For a data vector with component $x_i$ and a model vector $y_i$,  where $i=1,...,m$, the $\chi^2$ is defined as
\eqb 
\chi^2 = \sum_{i,j}^m\,[x_i-y_i] C_{ij}^{-1} [x_j-y_j].
\eqe 
In the above equation, $C_{ij}$ is the covariance matrix, estimated from $N$ independent realizations of mock data:
\eqb 
\hat{C}_{ij} = \frac{1}{N-1}\sum_{k=1}^N\,[x^k_i-\langle x^k_i \rangle][x^k_j-\langle x^k_j \rangle].
\label{eq: cov}
\eqe
Given a model with $p$ parameters, there are $m-p$ degrees of freedom in the $\chi^2$-fitting.
Due to the small number of mocks, we apply Jackknife re-sampling on the mocks by dividing each survey field into 18 sub-regions, giving a total of $N_J=54$ samples for each mock. The covariance matrix for an individual mock sample is estimated using equation~\ref{eq: cov}, with an extra factor $(N_J-1)$ to account for correlations between the Jackknife samples. We average over the covariance matrices of the 25 mocks to obtain the final covariance matrix. It is pointed out in \citet{jr:escoffier2016} that this method can reduce the noise on the covariance estimation, and fast approach the truth. However, we caution that these mocks are not completely independent, because they are constructed from the same N-body simulation~\citep{jr:gonzalez-perez}. 

The posterior of the model parameters $\theta$ given data $D$ is estimated in a Bayesian way:
\eqb 
P(\theta|D)=\frac{P(D|\theta)P(\theta)}{P(D)},
\eqe 
where $P(\theta)$ is the prior and $P(D)$ is treated as a normalization. The term $P(D|\theta)$ is proportional the likelihood $\mathcal{L}$, which we assume to be Gaussian:
$$\mathcal{L}\propto \exp(-\chi^2/2).$$ 
We use Monte-Carlo Markov Chain (MCMC) sampling of the parameter space, implementing the python package \texttt{emcee}\footnote{\url{http://dfm.io/emcee/}}.

\subsubsection{Data compression}

Instead of fitting the whole 2D correlation function, which requires an $N(r_p)\times N(\pi)$ -- dimensional covariance matrix, we compress the 2D information into the multipoles defined as
\eqb
\xi_{\ell}(r) = \frac{2\ell+1}{2}\int_{-1}^1\xi_c(r,\mu)\,\mathcal{P}_{\ell}(\mu)\,d\mu;\;\;\;\ell=0,2,4.
\label{eq: xi to multipoles}
\eqe
We ignore higher order multipoles because they are typically noisy and more sensitive to non-linearity. Multipoles are computed by interpolating the 2D correlation function, and this is done consistently for both the measurements and the models. In the QD model, we exclude $\xi_4$ from the fitting, because the non-zero signal at scales $r\geq10\mpcoh$ cannot be well reproduced by this model. 

We also considered adding the projected correlation function $w_p$:
\eqb
w_p(r_p)=\int_{-\pi_{\rm max}}^{\pi_{\rm max}}\xi(r_p,\pi)\,d\pi.
\label{eq: wp}
\eqe
This has the merit that it is in principle independent of RSD for large enough $\pi_{\rm max}$, and a knowledge of the true real-space clustering should be advantageous if we are focusing on redshift-space effects that cause deviations from this. However, we found in practice that it was not possible to choose a large enough $\pi_{\rm max}$ to achieve results that converged to the true real-space clustering without the results being too noisy to be useful. The $w_p$ statistic may be useful at small separations, $r<10\mpcoh$, as a means of probing the real-space 1-halo term, but as discussed above we do not need to do this in the present work.
We included $w_p$ in the fitting for the QD model; however, due to the limited information it could provide in addition to the multipoles, the $w_p$ was excluded in the HS model fitting.

For each of the six cross-correlation configurations, we fit the measurement simultaneously with its corresponding galaxy auto-correlation. This allows us to break the degeneracy between the galaxy and group bias.

\subsubsection{Scale cuts}

Below quasilinear scales $r\sim 10\mpcoh$, both models may fail to capture the full non-linear features. Fitting data points at these scales may introduce significant bias into the measured growth rate. Therefore, we test the models on a set of minimum fitting scales, $r_{\rm min}=2,5,10,15,20\mpcoh$ using the mock catalogues, and adopt the most appropriate cut for each subsample. For the HS model, because the model is designed to be able to fit smaller scales, we only test the model at $r_{\rm min}=2, 5, 10\mpcoh$.

\subsubsection{Integral constraints}
To account for the missing power for modes larger than the GAMA survey scale, we include the integral constraint $I$, which is a small constant added to the 2D correlation function. The expected integral constraint is given by:
\begin{equation}
    I\equiv\frac{b_1b_2}{3}\sigma_{\rm eff}^2=\frac{b_1b_2}{3}\int \Delta^2(k)\,W^2(k, r=r_{\rm eff})\, d\ln k,
    \label{eq: IC}
\end{equation}
where $\Delta^2(k)$ is the dimensionless linear matter power spectrum, $b_1$ and $b_2$ are the tracer biases, $W(k, r)=3[\sin(kr)/(kr)^3-\cos(kr)/(kr)^2]$, and $r_{\rm eff}$ is the effective radius of the survey volume in one of the GAMA fields, $V = 4\pi r_{\rm eff}^3/3$. The factor $1/3$ accounts for the fact that we combine measurements over three GAMA fields. Eq.~\ref{eq: IC} gives $I=0.0017b_1b_2$. This can also be measured in the mock data directly, by comparing the projected correlation function $w_p$ at large scales between the average of 25 samples and the combination of all mock samples. The measured values are consistent with the expectation given the statistical errors. In the QD model, we allowed $I$ to be a free parameter, and found that it has little impact on other parameters, with a posterior consistent with zero (e.g. see Table~\ref{tab: Model1 sum mock} and \ref{tab: Model1 sum data}). The integral constraints are then fixed to the measured values from mock for the streaming model, as shown in Table~\ref{tab: Model2 sum mock} and \ref{tab: Model2 sum gama}.

\subsubsection{Priors}

The parameters used in the two models and their uniform prior ranges can be found in Table~\ref{tab:prior}. For each of the group-galaxy subsample, the galaxy auto-correlation is fitted simultaneously with the cross-correlation. Parameters with subscript `$a$' are used for auto-correlations and `$c$' for cross-correlations.
There is another cosmological parameter that should be considered: the normalisation of the (linear) matter power spectrum $\sigma_8$. From Eq.~\ref{eq:galaxy_ps}, it is clear that on linear scales, $\sigma_8$ and $b$ are completely degenerate, hence RSD measurements are usually quoted in the combination $f\sigma_8$.
At large $k$, the shape of the non-linear power spectrum is actually sensitive to $\sigma_8$. However, such dependence is weak for the scales probed here, and we fix $\sigma_8=0.81$ throughout the analysis.

\begin{table}
	\centering
	\caption{Range of the uniform priors of the RSD fitting parameters. For growth rate, the usual constraint from RSD is $f\sigma_8$, but we fix $\sigma_8=0.81$ in this analysis. The $\alpha$, and $\eta$ parameters are the 1-halo parameters applied in the GSM model only.}
	\label{tab:prior}
	\begin{tabular}{lll}
		\hline
		Parameter & Prior (QD) & Prior (HS)\\
		\hline
		$b_{\rm gal}$ & $[0.1,2.5]$ & $[0.5,2]$ \\
		$b_{12}$  & $[0.1,2.5]$ & $[0.25,3]$ \\
		$f$       & $[0,2]$ & $[0,2]$ \\
		$\sigma_a$ [km\,s$^{-1}$]& $[143,1140]$ & $[30, 800]$\\
		$\sigma_c$ [km\,s$^{-1}$]& $[143,1140]$ & $[30, 800]$\\
		$I_a$     & $[0,0.1]$ & fixed \\
		$I_c$     & $[0,0.1]$ & fixed \\
		$\alpha_a$ & - & $[0.1,2]$ \\
		$\alpha_c$ & - & $[0.1,2]$ \\
		$\eta_a$ & - & $[0.5,2.5]$ \\
		$\eta_c$ & - & $[0.5,2.5]$ \\
		\hline
		
	\end{tabular}
\end{table}

\section{Results}
\label{sec:model-fitting}

\subsection{Mocks}
\label{sec:results-mocks}
\begin{figure*}
\centering
    \includegraphics[width=0.75\textwidth]{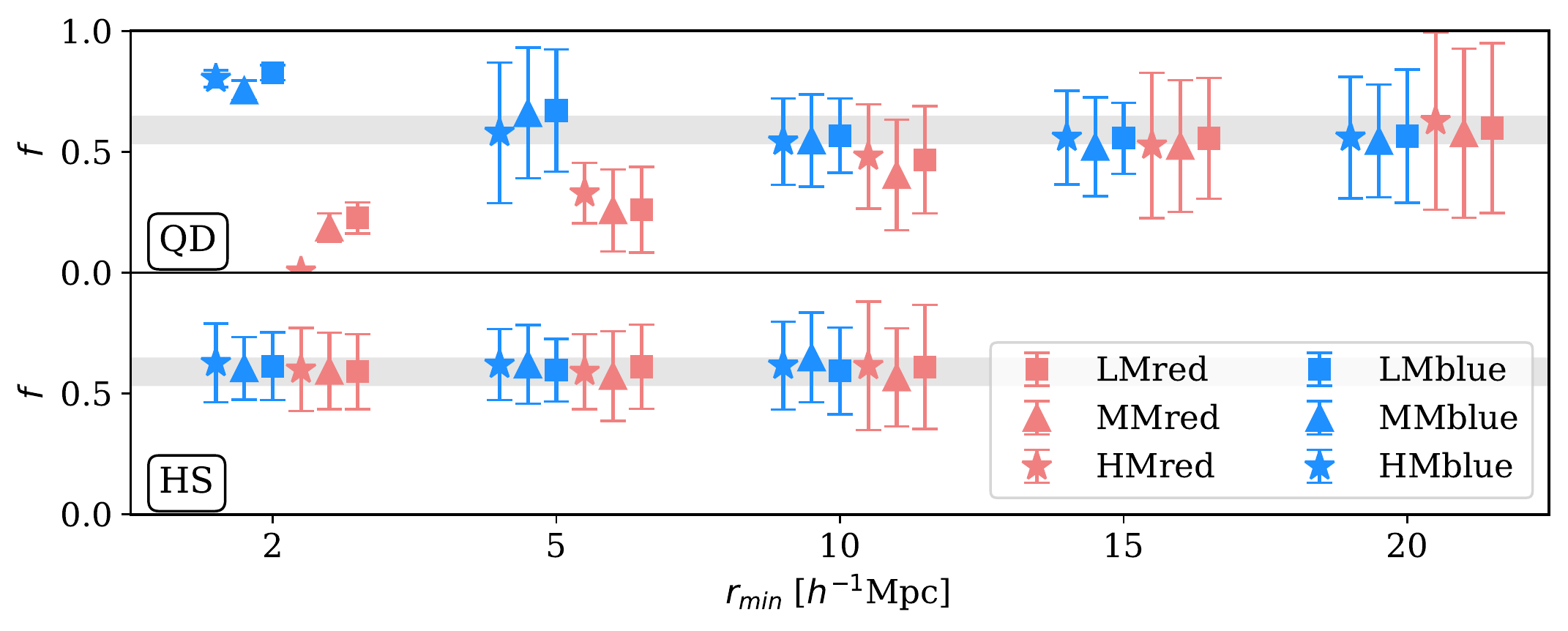}
    \caption{The means and scatter of the best-fit growth rate $f$ from 25 mocks as a function of the minimum fitting scale, $r_{\rm min}$, for the quasilinear dispersion (QD: top) and the halo streaming (HS: bottom) models. Data points at each $r_{\rm min}$ are displaced by $0.1\mpcoh$ for clarity. The grey band marks the $10\%$ regions around the mock fiducial value $f=0.593$ at $z=0.195$. Note that the error bars are for a single survey, so that the errors on the mean of the mocks are 5 times smaller than shown.}
    \label{fig: mock_params_M1}
\end{figure*}

We fit both models to each of the 25 mock samples, and compute the mean and scatter of the best-fit parameters. The aim is to assess the scale at which an unbiased growth rate can be recovered. The result is shown in Fig.~\ref{fig: mock_params_M1} for the set of $r_{\rm min}$ as mentioned in previous sections, and for each of the six configurations. The fiducial value of $f$ with $\pm 10\%$ range is marked by the grey band in each panel.
The error bar is comparable to, but should not be taken directly as the expected error size on the GAMA sample. 
The specific values of all model parameters are summarised in Tab.~\ref{tab: Model1 sum mock} and Tab.~\ref{tab: Model2 sum mock}.

Notice that in the case of halo streaming model, there is a caveat that the 1-halo templates are obtained from the average of the same set of mocks as they are tested on. Ideally, we would like to have access to multiple sets of simulations covering different cosmology and HOD prescription, with matched survey configurations as GAMA. Then, we would test the halo streaming model on by extracting the 1-halo templates from one set of simulations and apply it to the measurements from the others. In this way, we can assess whether the model is robust against bias due to a different cosmology or change of the simulation settings. Such test will be particularly relevant for the forth-coming large data sets, where the demand of the precision of the model is high. However, this is beyond the scope of this paper given the noise level of the GAMA data. We would like to defer such detailed comparison to a future study.

The top panel of Fig.~\ref{fig: mock_params_M1} shows the recovered growth rate $f$ using the QD model (Section~\ref{sec:model1}). 
As expected, when the small scales are included ($r_{\rm min}\leq5\mpcoh$), the fitted growth rates are significantly biased in all configurations, while at larger scales ($r_{\rm min}\geq15\mpcoh$), they converge to the fiducial value. 
The overall growth rate seems to be under-estimated by about $5-10\%$ for most scale cuts, but this is much smaller compared to the statistical error of the GAMA sample.
It is noticeable that the blue configurations are less biased down to smaller scales, with $f$ recovered to within 10\% at $r_{\rm min}=5-10\mpcoh$, compared to the red configurations which are only unbiased at $r_{\rm min}=15-20\mpcoh$. 
This may be due to the smaller FoG effect in the blue configurations compared to the red. 
From this test, we choose to adopt $r_{\rm min}=\{10,10,10,15,20,20\}\mpcoh$ for the LMblue, MMblue, HMblue, LMred, MMred, and HMred subsamples respectively for the QD model in the application to the GAMA data. 
The bottom panel of Fig.~\ref{fig: mock_params_M1} shows the the recovered growth rate $f$ using the HS model, where the results are impressively consistent. The growth rates for the different subsets are consistent to within an rms of 3\% in the mock average results, and the global average of these different subsets is within 2\% of the fiducial value. This successful performance holds down to even $r_{\rm min}=2\mpcoh$, although the estimated errors at that point are little different to those at $r_{\rm min}=5\mpcoh$, so we conservatively adopt the larger figure in our HS analysis.

\begin{figure*}
    \includegraphics[width=\textwidth]{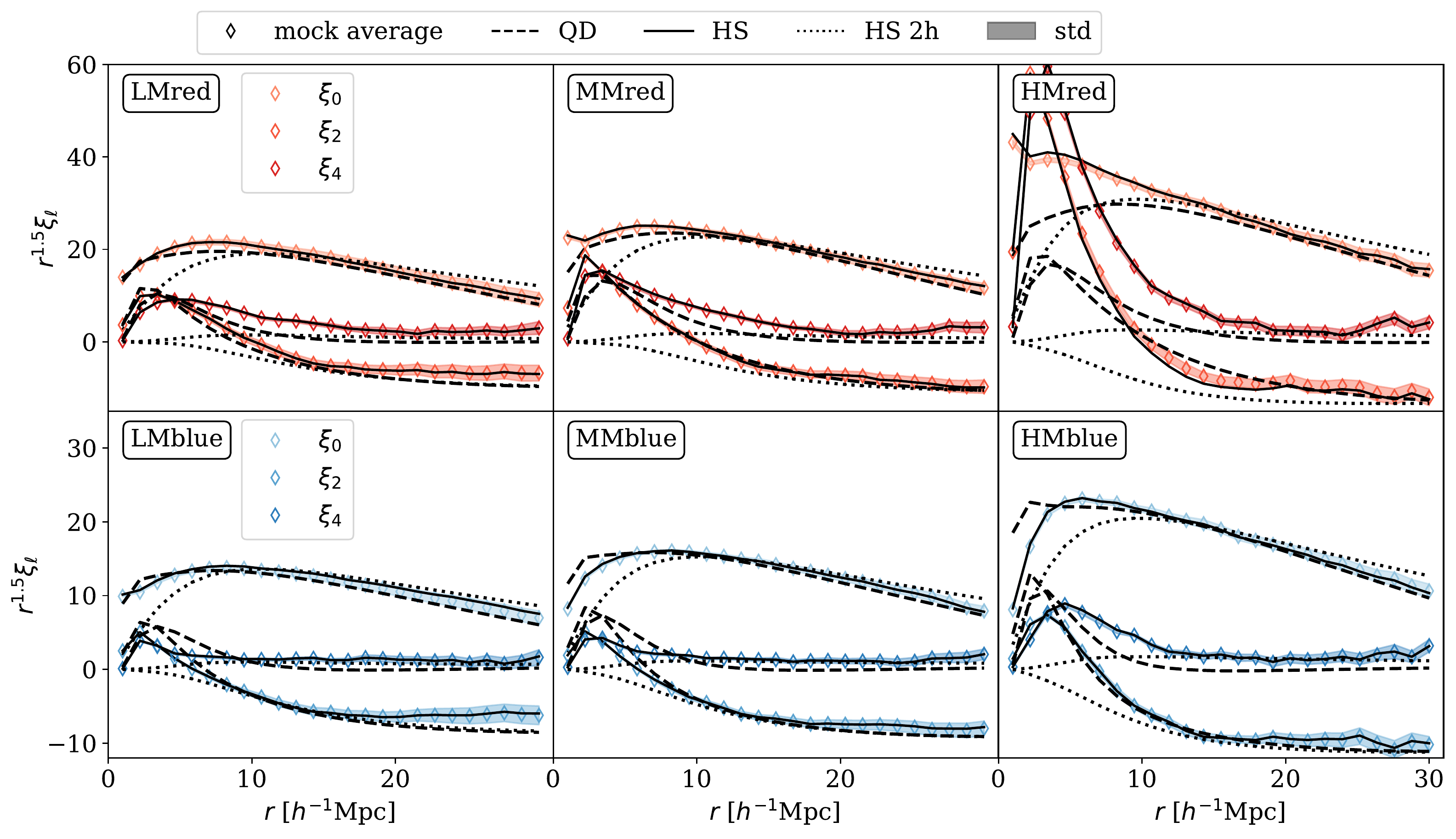}
    \caption{The multipoles of the group-galaxy cross-correlation functions in the mock average (open diamond). The coloured bands show the scatter on the mean. The best-fit QD models are shown in dashed black line, with $r_{\rm min}=\{10,10,10,15,20,20\}\mpcoh$ for the LMblue, MMblue, HMblue, LMred, MMred, and HMred subsamples respectively. The best-fit Gaussian streaming models with a 1-halo template are shown in solid black lines, with a fixed $r_{\rm min}=5\mpcoh$ for all sub-samples, with the corresponding 2-halo term shown in dotted black lines. For the presentation purpose, the multipoles have been multiplied by $r^{1.5}$.}
    \label{fig: wpxi02_mock_fittings_M1}
\end{figure*}

Fig.~\ref{fig: wpxi02_mock_fittings_M1} shows the linear and streaming model with the mean best-fit parameters from the mock subsamples, at the respective $r_{\rm min}$ as mentioned above. The mock average measurement as well as the $1\sigma$ error on the mean is also shown. In addition, we also show the corresponding 2-halo term of the streaming model in dotted black lines.
On large scales ($r\geq15\mpcoh$), all of the model curves converge, and match well with the mock average. It is noticeable that the full streaming model (with the addition of the 1-halo template) and its 2-halo term do not coincide exactly on these scales: the extracted 1-halo template still has some residuals in the monopole and quadrupole. The largest difference is seen in the hexadecapole. The slightly positive values seem to be produced only by the 1-halo FoG, which both the linear and the 2-halo terms of the streaming model fail to capture. Looking at smaller scales ($r\leq10\mpcoh$), it seems that the QD model under-predicts the power in the red configurations, and over-predicts that in the blue configurations.

\subsection{GAMA}
\label{sec:results-gama}

\begin{figure*}
    \includegraphics[width=\textwidth]{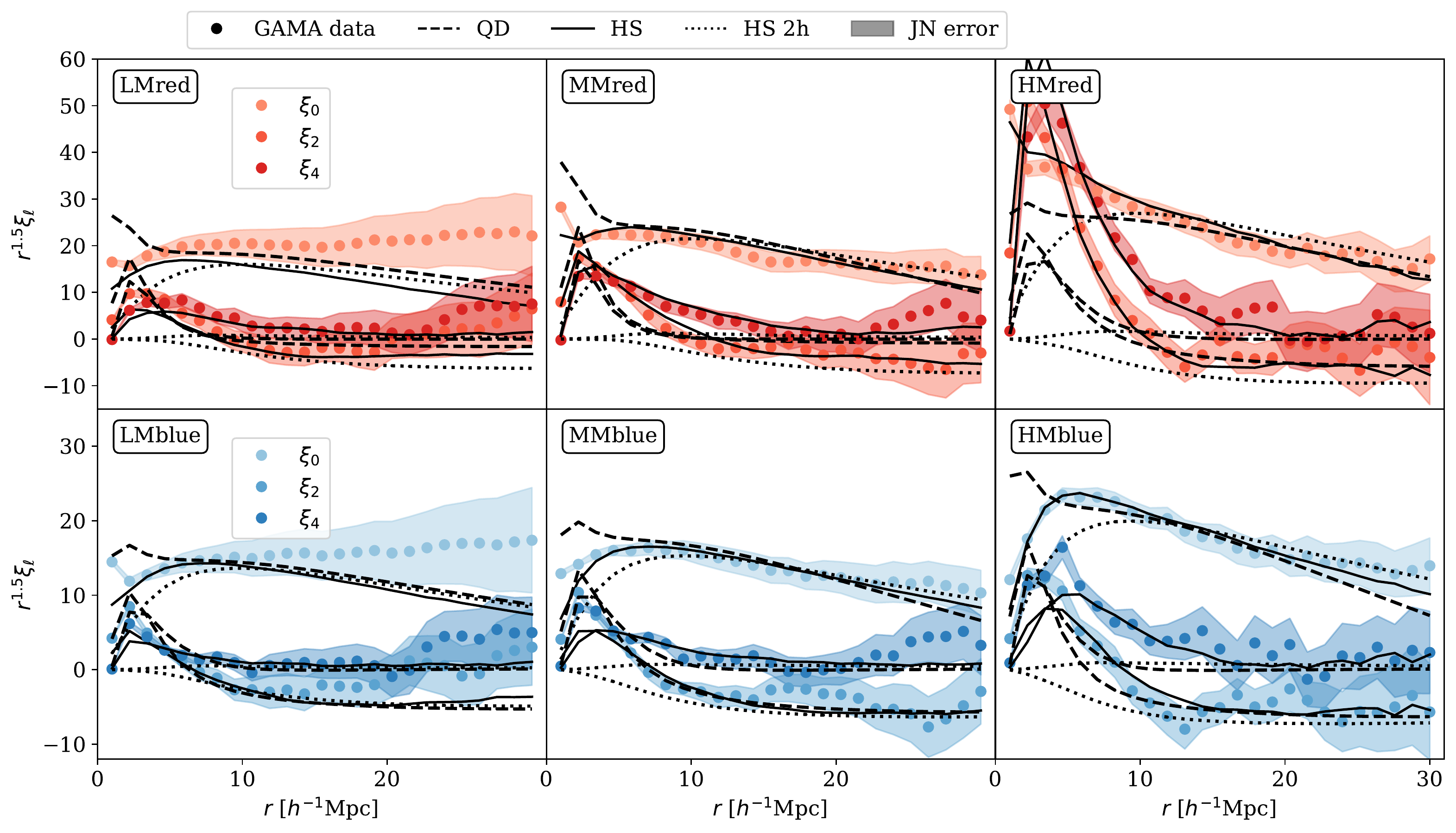}
    \caption{Same as \ref{fig: wpxi02_mock_fittings_M1} but for the actual GAMA data, with the same $r_{\rm min}$ adopted. The coloured bands show Jack-knife errors.  The $\chi^2$ for each of the models can be found in Table~\ref{tab: Model1 sum data} and \ref{tab: Model2 sum gama}.}
    \label{fig: wpxi02_GAMA_mock_fittings_M1}
\end{figure*}

\begin{figure*}
\centering
    \includegraphics[width=0.7\textwidth]{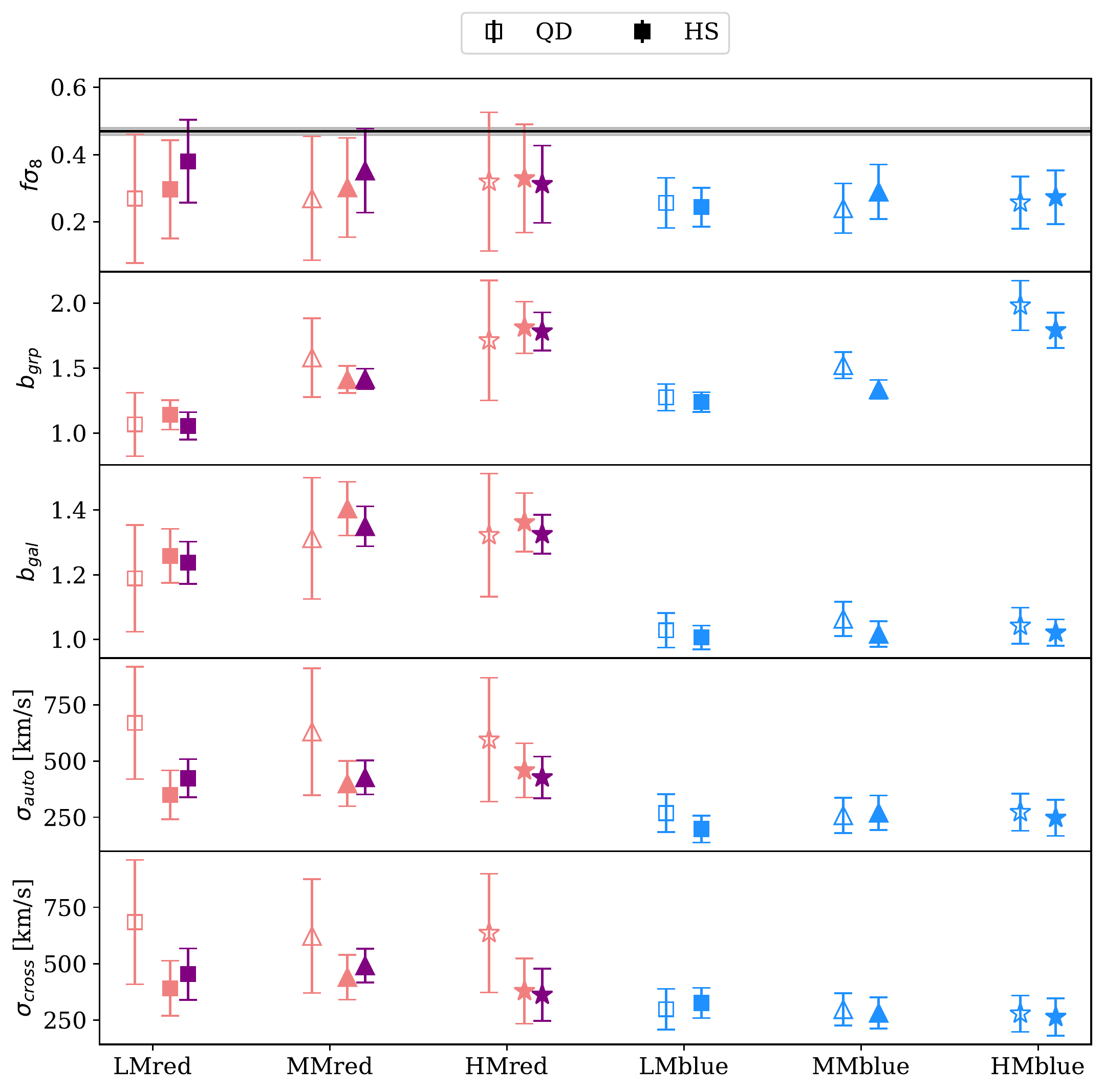}
    \caption{The MCMC parameter constraints for the actual GAMA data using the QD model (open symbols) and the streaming model with a 1-halo template (filled symbols). Each model is fitted with the respective $r_{\rm min}$ as in Fig.~\ref{fig: wpxi02_mock_fittings_M1}. The filled purple points show the constraints obtained using the 1-halo template from the `contaminated' red sample in the mock data. The 1-halo parameters are marginalised over in the HS model. 
    On the top panel, we have converted the constraints back to $f\sigma_8$ by multiplying back the fiducial $\sigma_8(z=0.195)$. 
    The black line and the grey band on the top panel mark $f\sigma_8=0.47\pm0.01$, the fiducial growth rate at $z=0.195$ using $\Omega_m$ and $\sigma_8$ constraints from \protect\cite{2018:planckpars}. The specific values for all model parameters are shown in Table~\ref{tab: Model1 sum data} and \ref{tab: Model2 sum gama}.}
    \label{fig: GAMA_params_M1}
\end{figure*}

Fig.~\ref{fig: wpxi02_GAMA_mock_fittings_M1} shows the measured GAMA multipoles (filled circles), the best-fit QD models (black dashed lines), and the HS models (black solid lines). In addition, the corresponding HS model 2-halo term is shown in the dotted black lines. The same scale cuts, $r_{\rm min}$, are adopted as in the mock case for each of the models. The $\chi^2$ and parameter values are shown in Table~\ref{tab: Model1 sum data} and \ref{tab: Model2 sum gama}. The full HS model 2D models are contrasted with the GAMA data in Appendix \ref{appendix:2dfits}.

We see that the QD model provides a reasonable fit to the monopole and quadrupole at given $r_{\rm min}$ in most configurations. The only exception is the LMred and LMblue subsamples, where the monopole power is boosted at large scales and the quadrupole power is consistent with zero. Despite the visual discrepancy, the $\chi^2$ of these models are consistent with the degrees of freedom of the data. The HS model is able to capture the shape of the multipoles down to smaller scales, especially the hexadecapole at scales $r>5\mpcoh$. At smaller scales ($r<5\mpcoh$), although excluded from fitting, the mock 1-halo template continues to provide a reasonable fit to the red configurations. But this is not the case for the blue configurations, where the non-linear velocity dispersion seems to be stronger in the actual data compared to the mock catalogues.
One possible explanation could be the impact of redshift measuring errors. These are not included in the mocks, and so any measured velocity dispersion in the real data only will include the redshift error in quadrature. The typical GAMA error is 50\,km\,$^{-1}$, but in detail  \citet{jr:gamaDR2} showed that redshift errors can depend on spectral and target properties. The redshift error for galaxies classified as the `absorption' type (i.e. the spectrum is dominated by absorption features) is $101$\,km\,s$^{-1}$,  compared to the `emission' type, which is $33$\,km\,s$^{-1}$. But red galaxies have a larger measured velocity dispersion, so the impact of redshift errors on the total measured dispersion will be less in that case.

Fig.~\ref{fig: GAMA_params_M1} shows the the mean and $1\sigma$ error on the model parameters from the MCMC posterior for the GAMA data, fitted at respective $r_{\rm min}$. The open and filled symbols denote parameter constraints from the QD model and the HS model respectively. In the latter case, we also show the constraints measured using the 1-halo template from the `contaminated' red galaxy sample (purple filled symbols). All sets of constraints show good consistency. 
Notice that the size of the error bar in the blue configurations is similar in both models, although the QD model has a scale cut at $10\mpcoh$ while the HS model at $5\mpcoh$. This is the consequence of the additional nuisance parameters added in the latter model.
The specific parameter values, including the 1-halo parameters in the HS model, can be found in Table~\ref{tab: Model1 sum data} and \ref{tab: Model2 sum gama}. In Fig.~\ref{fig: M1-autocolor-GAMA-blue-MCMC}-\ref{fig: GSM-autocolor-GAMA-blue-MCMC}, we further show the full posteriors from MCMC for all parameters in both models, grouped by the red and blue configurations. In the HS model, the 1-halo parameters $\alpha$ and $\eta$ have no primary degeneracy with the growth rate, although the growth rate can be shifted slightly through their small degeneracy with the velocity dispersion parameters. In practice, one would always marginalise over the 1-halo parameters. 

\begin{figure}
\centering
    \includegraphics[width=0.45\textwidth]{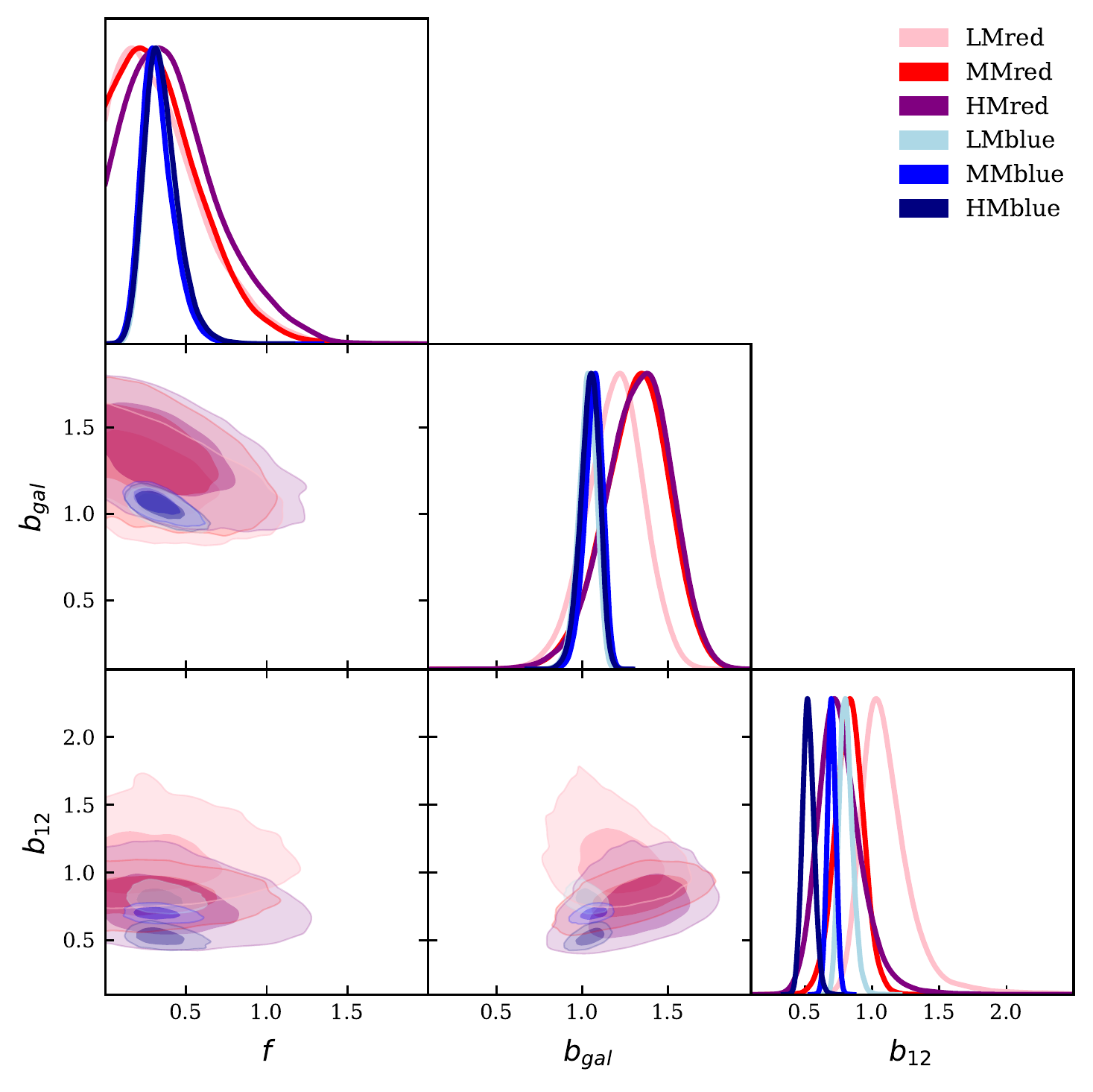}
    \caption{Marginalised MCMC posteriors for the QD model.}
    \label{fig: M1-autocolor-GAMA-MCMC-3pars}
\end{figure}

\begin{figure}
\centering
    \includegraphics[width=0.45\textwidth]{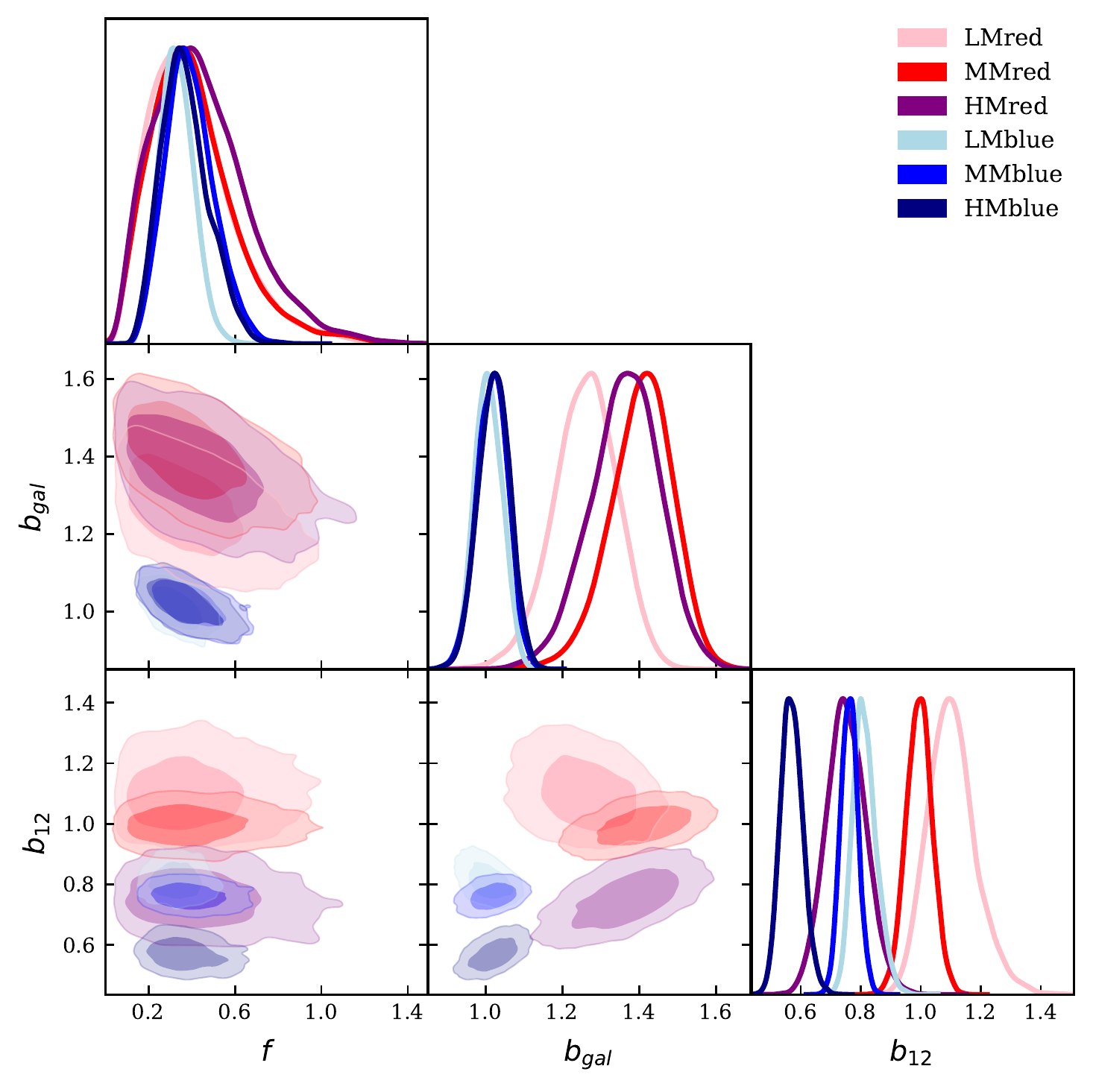}
    \caption{Marginalised MCMC posteriors for the HS model. The best-fit bias parameters are very different among different sub-samples, but the recovered $f$ values are consistent.}
    \label{fig: M2-autocolor-GAMA-MCMC-3pars}
\end{figure}

The middle two panels show the measured group and galaxy biases in both models. 
The LM, MM, and HM group biases measured consistently between the red and blue configurations in both models. 
For our selection of galaxies, we find that $b_{\rm gal}\approx1$ for the blue galaxies, and $b_{\rm gal}\approx1.3$ for the red galaxies; for the groups, we find that $b_{\rm grp}\approx 1.2, 1.4, 1.8$ for the low, medium and high mass ranges. 
The $b_{\rm grp}$ measurements are in qualitative agreement with that in \cite{2021MNRAS.506...21R} on large scales (e.g. see their Fig. 8), although direct comparison is non-trivial due to difference in the group selection.
The consistency between the two models is good in general, although for the blue configuration, the HS model gives systematically lower biases compared to the QD model by $0.5\sigma-1.5\sigma$.
The lower two panels show the measured auto- and cross-correlation velocity dispersion, $\sigma_a$ and $\sigma_c$. The two models measure consistent velocity dispersion, despite slightly different form for the FoG term. Notice that for the red configuration in the QD model, because of the large scale cut, the velocity dispersion posterior is prior driven.
There is a tentative trend (at $\sim 2\sigma$) that the red configurations have larger velocity dispersion compared to the blue configurations, with $\sigma_{a,c}\sim 400-500\, {\rm kms}^{-1}$ for the red configurations, and $\sigma_{a,c}\sim 300\, {\rm kms}^{-1}$ for the blue configurations in both auto- and cross-correlations. There is, however, no clear dependence on the group mass. 
These measured galaxy biases and velocity dispersion are in good agreement with other measurements from GAMA \citep[e.g.][]{jr:blake2013,jr:loveday}.
The marginalised posterior for $f$, $b_{\rm gal}$, and $b_{12}$ for the six configurations is shown in Fig.~\ref{fig: M1-autocolor-GAMA-MCMC-3pars} and \ref{fig: M2-autocolor-GAMA-MCMC-3pars} for the QD and HS models respectively.

The top panel shows the measured growth rate, consistent across the six subsamples for both models. Here, we have presented the results in the more general form of $f\sigma_8(z)$. The rationale for this is that our modelling assumes that the background cosmology (WMAP7 parameters) is known exactly. This is not precisely true, and the observed distortion parameter, $\beta=f/b$, is actually $\propto f\sigma_8$ (since $b\sigma_8$ is observable). We therefore multiply our fitted $f$ by the fiducial $\sigma_8(z)$ in order to obtain a combination that should be insensitive to the exact fiducial model.

We also note that RSD analyses commonly also allow for the Alcock-Paczynski effect \citep{1979Natur.281..358A, 1996MNRAS.282..877B}, which introduces additional distortions of the 2D correlation function from distance measurements using a `wrong' cosmology. This degree of freedom can boost the errors on $f\sigma_8$ substantially if the cosmological model is left free. But the AP effect is unimportant if the model is constrained by precise external CMB data as here. A further reason that this is reasonable is that the interest in RSD comes from the desire to test gravity: the CMB data give a precise prediction of $f\sigma_8$ and we want to know if this is what we measure.

In detail, then, we take $\sigma_8(z)=g(z)\sigma_8(0)=0.73$, where $z=0.20$, $\sigma_8(0)=0.81$, and $g(z)$ is the time-dependence of the (linear) density fluctuation in linear theory, normalised to $g(0)=1$. The measurements give a mean of $f\sigma_8=0.27$ with uncertainties ranging from $0.07-0.20$ for the QD model, and $f\sigma_8=0.29$ with uncertainties ranging from $0.07-0.14$ for the HS model. 
We combine our measurements from the six cross-correlation configurations for the HS model, accounting for their correlations using the mock catalogues. We compute the scatter on the average as well as the covariance of the best-fit $f$ for the six configurations in the 25 mock samples. Ideally, one would like to use the the full posterior. However, this would require the time consuming step of running MCMC for each of the mock sample, thus the simple average of the maximum likelihood values is adopted. Our combined constraint from the HS model thus gives
\begin{equation}
    f\sigma_8(z=0.20)=0.29\pm0.10.
\end{equation}
The corresponding figure for the QD model is $0.27\pm 0.14$, showing the extra information gained through the smaller scales that the HS model is able to probe.
We note that the limited number of mock samples means that our covariance matrices will be imprecise, so that the errors on the growth rate for an individual sample may be underestimated \citep{2007A&A...464..399H, 2016MNRAS.456L.132S}. However, the empirical dispersion in the mean of the maximum-likelihood values should be robust.

The striking thing about this GAMA-based figure is that it is rather low compared to the fiducial {\em Planck} figure of $f\sigma_8(z=0.20)=0.47\pm0.01$, derived from the {\em Planck} TT, TE, EE+lowE+lensing cosmological parameters \citep{2018:planckpars}:
our figure is $1.8\sigma$ below this {\em Planck} value.
This discrepancy is certainly unexpected given how well our modelling was able to account for the RSD signal in the different mock realizations, and how the recovered growth rates were consistent between different methods of model fitting. Furthermore, the figures recovered from the different GAMA subsamples show the same level of consistency with each other as is seen in subsamples within the mocks.

There are a number of things that can be said about the low observed figure. The first is that there is some evidence that the fiducial {\em Planck} figure may be too high, with local gravitational lensing data consistently arguing for a reduction of about 10\% (see e.g. \citejap{2021MNRAS.501.1481H}). Our measurement would then be in $1.3\sigma$ disagreement with a revised fiducial value of 0.42, implying that GAMA is an unusual dataset, but not unreasonably so. And we do have evidence that this is the case: inspection of Fig. \ref{fig: mock_gama_rand_zdist_0.1_0.3} shows that $N(z)$ has a substantial dip at $z\simeq0.24$, which is seen consistently in all three fields. One might suspect a problem with the redshift pipeline, but this feature is absent in a subsequent fourth GAMA field, not used here; the three main GAMA fields are simply rather unusual regions of space.
Finally, note that a multitracer analysis of RSD in GAMA by \citet{jr:blake2013} gave $f\sigma_8(z=0.18)=0.36\pm0.09$, which is also slightly lower than {\em Planck}, albeit not inconsistently so.

\subsection{Group bias}
\label{sec: group bias}

\begin{table*}
	\centering
	\caption{Bias for the groups in the LM, MM, and HM stellar mass bins for the mock average and GAMA. The first column shows the mean bias value computed from the fitting formula \protect\cite{2010ApJ...724..878T}, and the second column shows the corresponding bias computed at the mean halo mass in each case. The next two columns marked with a `*' show the bias computed in the same way, but with the GAMA group halo mass computed from a more up-to-date mass-luminosity relation \citep{2022MNRAS.510.5408R}. The rest of the columns show the fitted biases from the six cross-correlation configurations in the mocks and the GAMA data.}
	\label{tab:biases}
	\begin{tabular}{llcccccccc}
		\hline
		Group bias & &  T10 & T10 $b(\bar{M}_h)$ &T10$^*$ & T10$^*$ $b(\bar{M}_h)$ & QD-red & QD-blue & HS-red & HS-blue\\
		\hline
		Mocks & LM & $1.02$ & $0.92$ & - & - & $1.20\pm0.04$ & $1.20\pm0.02$ & $1.18\pm0.03$ & $1.20\pm0.02$\\
		& MM & $1.26$ & $1.13$ & - & - & $1.48\pm0.05$ & $1.46\pm0.02$ & $1.42\pm0.02$ & $1.38\pm0.02$\\
		& HM & $1.83$ & $1.65$ & - & - & $1.90\pm0.06$ & $2.09\pm0.03$ & $1.96\pm0.03$ & $1.92\pm0.03$\\
		\hline
        GAMA & LM & $1.00$ & $0.96$ & 1.12 & 1.11 & $1.07\pm0.24$ & $1.27\pm0.10$ & $1.14\pm0.11$ & $1.24\pm0.08$\\
        & MM & $1.20$ & $1.17$ & 1.41 & 1.39 & $1.58\pm0.30$ & $1.52\pm0.10$ & $1.41\pm0.10$ & $1.34\pm0.07$\\
        & HM & $1.52$ & $1.49$ & 1.85 & 1.81 & $1.71\pm0.46$ & $1.98\pm0.19$ & $1.81\pm0.20$ & $1.79\pm0.14$\\
		\hline
	\end{tabular}
\end{table*}

Finally, it is interesting to ask if the group biases that we measure are in accord with what is expected for haloes of these masses. We compute the expected group bias in GAMA from the calibrated halo mass for the groups based on Eq.~\ref{eq: mh vs L}.
We adopt the \citet{2010ApJ...724..878T} fitting formula for the linear halo bias. The halo bias is expressed in terms of the peak height parameter $\nu\equiv\delta_c/\sigma_R$, where $\delta_c\approx1.686$, and $\sigma_R$ is the rms of the linear power spectrum filtered with a spherical top hat function with radius $R$ (cf. Eq.~\ref{eq: IC}).
This is related to the halo mass via $M_h=200\bar{\rho}_m 4\pi R^3/3$, where $\bar{\rho}_m$ is the mean background density.
The mean group bias in a stellar mass range is estimated by
\begin{equation}
\hat{b}_{\rm grp}=\frac{\sum_i N(\log M^i_h)\,b(\log M^i_h)}{\sum_i N(\log M^i_h)},
\label{eq: bias}
\end{equation}
where $N(\log M^i_h)$ is the number of groups in the logarithmic halo mass bin $i$. The halo mass adopted in case of mock and GAMA are as shown in Fig.~\ref{fig: stellar_lensing_mass}.
For GAMA, we also include the uncertainty in the calibrated halo mass due to the uncertainties in $\log M_p$ and $\alpha$ in Eq.~\ref{eq: mh vs L}. We combine the error via:
\begin{equation}
\Delta_{\log M_h}=\Delta_{\log M_p} + |\Delta_{\alpha}\log(L_{\rm grp}/L_0)|.
\end{equation}
For the mass range concerned here, $\Delta_{\log M_h}=0.8-0.2$ dex from low mass to high mass. We account for this scatter by convolving the number of objects with a Gaussian distribution with width $\Delta_{\log M_h}$ in each $\log M_h$ bin. 
The predicted and measured group biases using mocks and GAMA data are summarised in Table~\ref{tab:biases}. We also show biases computed at the logarithmic mean halo mass $\bar{M}_h$.
These values are close to the average bias computed from Eq.~\ref{eq: bias} in the case of GAMA, but they deviate from the mock average significantly. This indicates that estimating the bias from the mean halo mass depends heavily on the distribution of the halo mass of the sample considered.
In addition, we include the case where the more up-to-date mass-luminosity relation from \citet{2022MNRAS.510.5408R} is used to compute the group halo mass. The equivalent parameters to Eq.~\ref{eq: mh vs L} are $M_p=(8.1\pm0.4)\times10^{13}h^{-1}M_{\odot}$, $L_0=10^{11.5}h^{-2}L_{\odot}$, and $\alpha=1.01\pm0.07$. The halo mass computed with this calibration is larger than using the fiducial \cite{jr:han}, resulting in good consistency with the fitted group bias from both the QD and HS models as shown in Table~\ref{tab:biases}.

In the mock catalogues, the predicted group bias for the LM, MM, and HM subsamples are all lower than the fitted values.
These differences are significant given the error bar on the average of the mock measurements. In all cases, the group bias has apparently been under-estimated by about $20\%$.
This deviation in the group bias may arise because the arithmetic mean host halo mass of the group members is used as a proxy for the group halo mass. However, if one uses the total mass of unique host haloes in the group as $M_h$, then the bias in each mass range only increases by $\sim10\%$. 
Since the mock group masses are calibrated using `real' simulation halo masses, the difference illustrates clearly that our galaxy groups are not in 1-to-1 correspondence with single haloes, emphasising once again the importance of analysing real and mock data with the same group finder.

\section{Summary and Conclusions}
\label{sec:conclusions}

In this work, we have investigated the Redshift-Space Distortions (RSD) of group-galaxy cross-correlations, with the aim of understanding the robustness with which measurements of the density fluctuation growth rate can be extracted from such measurements. We have focused on the differences in the measured RSD using different types of galaxy and group, and developed new methods for fitting such data down to the small-scale nonlinear regime.

We have used data from the GAMA survey in the redshift range $0.1<z<0.3$ to measure the 2D cross-correlation function $\xi(r_p,\pi)$ between groups and galaxies. The groups were found using a FoF algorithm from \cite{jr:katarina1}, and were subdivided into three stellar mass bins (LM: 40\%, MM: 50\%, and HM: 10\%). The corresponding halo mass for the groups was calibrated using the relation in \cite{jr:han}, and the groups are expected to have typical masses of $(10^{12.2}, 10^{12.7}, 10^{13.2})\, M_\odot$. 
For \citet{2022MNRAS.510.5408R}, the mean halo masses are $(10^{12.6}, 10^{13.1}, 10^{13.5})\, M_\odot$.
The galaxies were split into red and blue subsets using a cut in the $g-i$ vs $z$ plane, yielding in total six cross-correlation configurations: LMred, MMred, HMred, LMblue, MMblue, and HMblue. 

We have used 25 GAMA lightcone mocks from \cite{jr:farrow2015} to test RSD models and to construct Jackknife covariance matrices for likelihood fitting. Mock group catalogues were generated using the identical algorithm that was applied to the real GAMA data.  The mock catalogues are distinct from observation in several aspects: the mean redshift distribution, the bimodal $g-i$ colour distribution, and the total stellar mass of the groups. We discuss the appropriate empirical selection that yields the best match between the mocks and the data subsamples.
The measured 2D correlation functions show good consistency between the data and the mocks down to small scales, and the same variation of the signals with galaxy colours and group masses are observed.
The different cross-correlation results yield group biases that increase with mass, as expected. For GAMA, the predicted group bias from \citet{2010ApJ...724..878T} is lower but consistent with the fitted values using the halo mass calibration in \citet{jr:han}, whereas that from \citet{2022MNRAS.510.5408R} agrees well with the fitted values. For mocks, however, these values tend to be higher than predicted. This difference illustrates that the groups found in redshift space do not constitute a pure halo sample.

We have compared these measurements with two RSD models: (1) a Quasilinear Dispersion (QD) model; (2) a novel Halo Streaming (HS) model.
The QD model is a generalization of the linear dispersion model of \cite{jr:mohammad} to use the
non-linear real-space power spectrum. We found from testing on the mocks that this model provides unbiased measurements of the growth rate at $r_{\rm min}=10-20\mpcoh$ depending on the subsample. The HD model uses a halo model decomposition of the correlations, where a streaming model 2-halo term is combined with an empirical 1-halo template adopted from the mock average. This promising model, with the addition of two nuisance parameters, allows unbiased results on the growth rate down to $r_{\rm min}=5\mpcoh$ when fitting individual mock realizations, and for all group-galaxy combinations. 
For the GAMA measurements, an MCMC analysis was used to obtain the posterior of our model parameters. 
We found that given the scale cuts, all of the subsamples recover consistent growth rates in both models. The average growth rate from the six subsamples using the HS model is $f\sigma_8=0.29\pm0.10$ at $z=0.20$, where the error should be robust as it is taken directly from the dispersion in maximum-likelihood values for the mock data. This figure is $1.8\sigma$ lower than the fiducial {\em Planck} value of $f\sigma_8=0.47\pm0.01$, and we have considered the implications of this result. At face value, the low GAMA result is consistent with the suggestions from gravitational lensing that the true value of $f\sigma_8$ may be about 10\% lower than the {\em Planck} central figure (e.g. \citejap{2021MNRAS.501.1481H}. But there are objective reasons to believe that the GAMA dataset may be a statistical outlier, based on known anomalies in the redshift distribution in the GAMA fields.

Therefore, the real test of the RSD modelling presented here will be when it can be applied to much larger and more precise datasets, such as the Bright Galaxy Sample from the Dark Energy Spectroscopic Instrument
(DESI) survey\\
\citep{2018SPIE10702E..1FM} and the the Wide Area VISTA Extra-Galactic Survey (WAVES) \citep[][]{2016ASSP...42..205D}.
We are greatly encouraged by the success of our halo streaming model in reproducing mock cross-correlations down to the smallest scales, and in yielding consistent values of $f\sigma_8$ from different tracers, to a tolerance of better than 3\%.
This hybrid approach, taking advantage of ever more realistic mock data, therefore seems an attractive way of obtaining robust constraints on the growth of cosmological density fluctuations, and we look forward to seeing it applied to next-generation surveys.

\section*{Acknowledgements}

QH was supported by Edinburgh Global Research Scholarship and the Higgs Scholarship from Edinburgh University. JAP and SA were supported by the European Research Council under the COSFORM grant no. 670193. YC acknowledges the support of the Royal Society through a University Research Fellowship and an Enhancement Award. YC thanks the hospitality of the Astrophysics and Theoretical Physics groups of the Department of Physics at the Norwegian University of Science and Technology during his visit. MB is supported by the Polish National Science Center through grants no. 2020/38/E/ST9/00395, 2018/30/E/ST9/00698, 2018/31/G/ST9/03388 and 2020/39/B/ST9/03494, and by the Polish Ministry of Science and Higher Education through grant DIR/WK/2018/12.

GAMA is a joint European-Australasian project based around a spectroscopic campaign using the Anglo-Australian Telescope. The GAMA input catalogue is based on data taken from the Sloan Digital Sky Survey and the UKIRT Infrared Deep Sky Survey. Complementary imaging of the GAMA regions is being obtained by a number of independent survey programmes including GALEX MIS, VST KiDS, VISTA VIKING, WISE, Herschel-ATLAS, GMRT and ASKAP providing UV to radio coverage. GAMA is funded by the STFC (UK), the ARC (Australia), the AAO, and the participating institutions. The GAMA website is \url{http://www.gama-survey.org/}.

\section*{Data Availability}
All of the GAMA data are publicly available on \url{http://www.gama-survey.org/dr3/schema/}. The mock catalogues and the GAMA group catalogue are available upon reasonable request.




\bibliographystyle{mnras}
\bibliography{references} 





\appendix

\section{The linear model for redshift-space cross-correlations}
\label{appendix: model linear}

The redshift-space 2-point correlation function can be expanded in terms of Legendre polynomials $\mathcal{P}_\ell(\mu)$ with $\ell=0,2,4$ \citep{jr:hamilton}:
\eqb 
\xi^s_g=\xi_0(r)\mathcal{P}_0(\mu)+\xi_2(r)\mathcal{P}_2(\mu)+\xi_4(r)\mathcal{P}_4(\mu).
\eqe 
The multipoles are given by
\begin{align}
\xi_0 (r) &=\Big(1+\frac{2}{3}\beta+\frac{1}{5}\beta^2\Big) \, \xi_g(r), \label{eq: xi0}\\
\xi_2 (r) &=\Big(\frac{4}{3}\beta+\frac{4}{7}\beta^2\Big) \, [\xi_g(r)-\bar{\xi}_g(r)],\label{eq: xi2}\\
\xi_4 (r) &=\frac{8}{35} \, \beta^2 \,\Big[\xi_g(r)+\frac{5}{2}\,\bar{\xi}_g(r)-\frac{7}{2}\,\bar{\bar{\xi}}_g(r)\Big], \label{eq: xi4}
\end{align}
where, 
\begin{align}
\xi(r)&=\int_0^\infty d\ln k \, \Delta^2(k) \, \frac{\sin(kr)}{kr}, \label{eq: xir}\\
\xib(r) &= (3/r^3) \int_0^r \xi(x)\, x^2\, dx \cr
&=\int {3\Delta^2(k)\over (kr)^3} \, d\ln k\, 
(\sin kr - kr\, \cos kr)\;; \label{eq: xirb}\\
\xibb(r) &= (5/r^5) \int_0^r \xi(x)\, x^4\, dx \cr
&=\int {5\Delta^2(k)\over (kr)^5} \, d\ln k\, 
(kr(6-k^2r^2)\cos kr - (6-3k^2r^2)\sin kr),\label{eq: xirbb} \cr
\end{align}
and the subscript `$g$' indicates an extra factor of $b^2$.
For practical purposes it is useful to have an explicit direct expression:
\eqb
\eqalign{
\xi_g^s(r,\mu) &= \xi_g + \left({2\beta\over 3} + {\beta^2\over 2}\right)\,\xib_g
-{3\beta^2\over 10}\, \xibb_g \cr
&+ \left(2\beta(\xi_g-\xib_g) + 3\beta^2(\xibb_g-\xib_g)\right)\, \mu^2 \cr
&+ \beta^2\left(\xi_g+5\xib_g/2 - 7\xibb_g/2\right)\, \mu^4.
}
\eqe

Since this
expression is linear in the different contributions to $P$, we immediately see that
Hamilton's expression can be generalised by the replacements $2\beta\to(\beta_1+\beta_2)$
and $\beta^2\to\beta_1\beta_2$.
In the case of two tracers with biases $b_{\rm gal}$ and $b_{\rm grp}=b_{\rm gal}/b_{12}$, equations~\ref{eq: xi0}~--~\ref{eq: xi4} are therefore modified to \citep{jr:mohammad} 
\begin{align}
\xi_{0,c}(r) &= \Big(1+\frac{1}{3}\beta_{\rm gal}(1+b_{12})+\frac{1}{5}\beta^2_{\rm gal}b_{12}\Big)\xi_c(r),\\
\xi_{2,c}(r) &= \Big(\frac{2}{3}\beta_{\rm gal}(1+b_{12})+\frac{4}{7}\beta^2_{\rm gal}b_{12}\Big)[\xi_c(r)-\bar{\xi}_c(r)],\\
\xi_{4,c}(r) &= \frac{8}{35}\beta^2_{\rm gal}b_{12}\Big[\xi_c(r)+\frac{5}{2}\bar{\xi}_c(r)-\frac{7}{2}\bar{\bar{\xi}}_c(r)\Big],
\end{align}
where $\beta_{\rm gal}=f/b_{\rm gal}$, and $\beta_{\rm grp}=fb_{12}/b_{\rm gal}$.  The various moments with a subscript `$c$' are defined as in equations~\ref{eq: xir}~--~\ref{eq: xirbb}, but with a factor $b^2_{\rm gal}/b_{12}$.

\section{The streaming model for redshift-space cross-correlations}
\label{appendix: model}

Redshift-space correlations are frequently discussed using the streaming model (e.g. \citejap{Fisher1995}), in which the total observed pair counts are related to the true counts in real space by a scale-dependent
shift and smoothing:
\eqb
1+\xi_s(\sigma,x) = \int \left(1+\xi_r(\sigma,y)\right)\, {\cal P}(x-y\mid y)\, dy,
\eqe
where $\sigma$ is the transverse separation and $x$ \& $y$ are the radial separations
in respectively redshift space and real space. For the correct true choice of the
distribution function $\cal P$, this equation should be exact by definition as it
just expresses conservation of pairs. The same equation should apply to both the
auto- and cross-correlation case.

Commonly, the distribution function is assumed to be a Gaussian:
\eqb
{\cal P}(x-y\mid y) = {1\over \sqrt{2\pi}\, \sigma(y)}\,
\exp\left(
-{
(x-y-u(y))^2\over 2\sigma^2(y)
}
\right),
\eqe
where $u(y)$ is the line of sight comoving offset caused by the mean
pairwise streaming velocity: $u(y)=\mu v_{12}(r)/aH$, where
$\mu=y/r$ and $r^2=\sigma^2+y^2$.  
Although we do not need to assume that the correlations are
small, it is convenient to adopt the standard linear-theory prescription for the mean pairwise velocity that results from a given correlation, in which
the mean proper relative infall velocity of
a pair of biased tracers at comoving separation $r$ is
\eqb
{v_{12}(r) \over aH r} =  - {1\over 3}\, f (b_1+b_2) \, \xib_m(r)
=  - {1\over 3}\, (\beta_1+\beta_2) \, \xib_x(r),
\eqe
where the cross-correlation for the tracers is related to the autocorrelation of the mass
by $\xib_\times = b_1b_2\xib_m$. The appearance of only $\beta$ factors without any explicit
bias parameter then makes it plausible that direct consideration of pairwise peculiar velocities
will yield the correct Kaiser--Hamilton form in the linear $\xi\ll1$ limit,
although proving this is not straightforward (see \citejap{Fisher1995}).

To achieve the correct linear limit, it is necessary to include not only
the mean streaming, but also the scale-dependent dispersion in pairwise velocity
that must arise in a Gaussian velocity field. 
Note that the velocity field smooths
$1+\xi$, and any effects on $\xi$ itself from the smoothing will introduce
corrections beyond linear order in $\xi$. Therefore, almost paradoxically, the linear
corrections to $\xi$ come from smoothing the unity term in $1+\xi$: this
gives a non-unity result because the smoothing is not exactly a convolution.
Therefore we have
\eqb
1+\xi_s(\sigma,x) = \xi_r(\sigma,x) + \int {\cal P}(x-y\mid y)\, dy.
\eqe
Incidentally, this shows that there is an error in Fisher's final equation
(26), which omits the critical unity term.

To evaluate the integral for the linear correction term, one can
Taylor expand the integrand.
But this needs to be done with care, since
the argument of the exponential will not always be small. 
We have to consider the effects of the scale-dependent streaming, $u(y)$,
and the scale-dependent squared dispersion, $V(y)\equiv \sigma^2(y)$. But both of these
quantities scale in proportion to $\xi$, so to linear order we will never
need to consider terms in which they are multiplied. Therefore, the two effects
can be considered in turn. Variations in $u$ 
so the leading effect is to scale $z$ by a factor $1+u'$ (where $u'\equiv du/dy$),
lead to a correction to $\xi$ of
\eqb
\Delta\xi = -u',
\eqe
where $u'\equiv du/dy$.
This is just a Jacobian correction.

Allowing for variation in the dispersion is more complicated. We need
to expand ${\cal P}$ up to order $z^4$, where $z\equiv y-x+u(x)$,
and use the Gaussian result 
that $\langle z^4 \rangle=3\langle z^2 \rangle^2$. This then yields
\eqb
\Delta\xi = V''/2,
\eqe
in agreement with Fisher's equation (24). This term is critical, as it generates
the hexadecapole distortion.

\bigbreak
\noindent
{\bf The pairwise dispersion}

\noindent
Implementing the streaming model thus requires two steps: (1) allowing
for the mean infall as a function of scale; (2) smoothing by a scale-dependent
pairwise dispersion, in order to achieve the linear Kaiser limit correctly.
The second step sounds like the FoG convolution, but that would be a separate correction that allows for a random `dither' in the large-scale velocity field. The smoothing here is present purely through the statistics of the linear velocity field. For the
streaming model, we require the line of sight pairwise velocity dispersion:
\eqb
\sigma^2 = \langle (v_z^b-v_z^a)^2 \rangle - \langle (v_z^b-v_z^a) \rangle^2.
\eqe
Recall that the averages here are density weighted, otherwise the
mean relative velocity would vanish. But the result of this is that 
$\langle (v_z^b-v_z^a) \rangle$ is of order the linear correlation
function, so $\langle (v_z^b-v_z^a) \rangle^2$ is of second order
in $\xi_{\rm lin}$. If we keep only the lowest order terms, then
\eqb
\sigma^2 = 2 \langle v_z^2 \rangle - 2\langle v_z^a v_z^b \rangle 
\eqe
(and the effects of density weighting are neglected as also being of higher order).
These terms can be evaluated using the theory of correlated velocity fields
as presented in \cite{Gorski1988}:
\eqb
\sigma^2 = 2\Psi_\perp(0) - 2\Psi_\perp(r) -2[\Psi_\parallel(r)-\Psi_\perp(r)]\,\mu^2,
\eqe
where $\mu = r_\parallel/r$ and
\eqb
\Psi_{\parallel,\perp}(r) = a^2H(z)^2f^2(z) \int \Delta^2(k,z)\, k^{-2}\, d\ln k\, K(kr),
\eqe
with the kernels being respectively
\eqb
\eqalign{
K_\perp(x) &= [\sin(x)-x\cos(x)]/x^3; \cr
K_\parallel(x) &= \sin(x)/x - 2[\sin(x)-x\cos(x)]/x^3,
}
\eqe
both of which tend to 1/3 as $x\to 0$.

\bigbreak
\noindent
{\bf A hybrid streaming + halo model}

\noindent
In real space, it is common to decompose the matter power spectrum as
\eqb
P(k)=P_{2h}(k) + P_{1h}(k),
\eqe
Where the 2-halo term is close to linear theory and the 1-halo term is
a discreteness term that reflects the finite number density of
haloes -- of the form of shot noise but filtered on small scales
to reflect the extended nature of haloes. Hang et al. (2021) proposed 
an effective model for decomposing the auto-power of biased tracers:
\eqb
P(k)=b_{2h}^2P_{2h}(k) + b_{1h}^2P_{1h}(k),
\eqe
where $P_{2h}$ and $P_{1h}$ are provided by HALOFIT.
In the same spirit, then, one might propose a model for cross-correlations:
\eqb
\xi_{12}=b_1b_2\xi_{2h} + b_{1h}^2 \xi_{1h}.
\eqe
It is not clear in advance that the 1-halo term is appropriate here, since
$\xi_{1h}$ is defined for autocorrelation and averaging over all halo masses,
so the appropriate shape of the 1-halo correction for cross-correlation
and with halo mass bins may not be universal. But this simple 1-parameter
form for the 1-halo term seems worth investigating as a first approximation.
More generally, we can add an empirical 1-halo term, $\xi_{12,1h}$, to the
large-scale linear power, which could be described via a number of
nuisance parameters: the 1-parameter $b_{1h}$ prescription is the simplest, but
one might have two parameters describing the slope and amplitude of a power law etc.

To convert to redshift space, we first have to allow for large-scale
peculiar velocities affecting halo positions, so that $b_1b_2\xi_{2h}$ is
replaced by the quasilinear $\xi_{12,s}$ discussed above:
\eqb
\xi_{12,s} = \xi_{12,{\rm stream}}(r_\perp,r_\parallel) + \xi_{12,1h}(r)
\eqe
The advantage of using the streaming model over linear theory is that the 
function includes non-trivial quasilinear corrections even when the linear
coupling between density and velocity is assumed. We can go further than this and
add an assumed quasilinear correction to the pairwise infall itself, by
making the reasonable assumption that the peculiar velocities will approximately
cancel the Hubble flow if the clustering is sufficiently nonlinear. This motivates the form
\eqb
{v_{12}(r) \over aH r} = 
- { (\beta_1+\beta_2) \, \xib_\times(r)/3 \over
1 + (\beta_1+\beta_2) \, \xib_\times(r)/3 }.
\eqe

Finally, we need to allow for nonlinear
peculiar velocities that convolve the galaxy distribution in the radial direction,
so that we convolve the correlation in $r_\parallel$ by some FoG damping factor, $D(\Delta r_\parallel)$:
\eqb
\xi_{12,s} \to \xi_{12,s} * D(\Delta r_\parallel).
\eqe
The FoG term is generally taken to be an exponential,
\eqb
D(\Delta r_\parallel)=(\sqrt{2}\sigma_{\rm FoG})^{-1}\exp(-\sqrt{2}|\Delta r_\parallel|/\sigma_{\rm FoG}),
\eqe
corresponding to a Lorentzian filtering of the power spectrum:
\eqb
{\tilde D}(k_\parallel) = 
\left(1 + (k_\parallel \sigma_{\rm FoG})^2/2 \right)^{-1}.
\eqe
As we argue in the body of the paper, it is necessary to use a different convolution for the 1-halo and 2-halo terms. But in any case we have not derived a full model for the 1-halo term in redshift space, and in the present paper we take this empirically from simulations, supplementing the above 2-halo term described via the streaming model.

\section{Fits to individual datasets}
\label{appendix:2dfits}

\begin{figure*}
\centering
\includegraphics[width=0.8\textwidth]{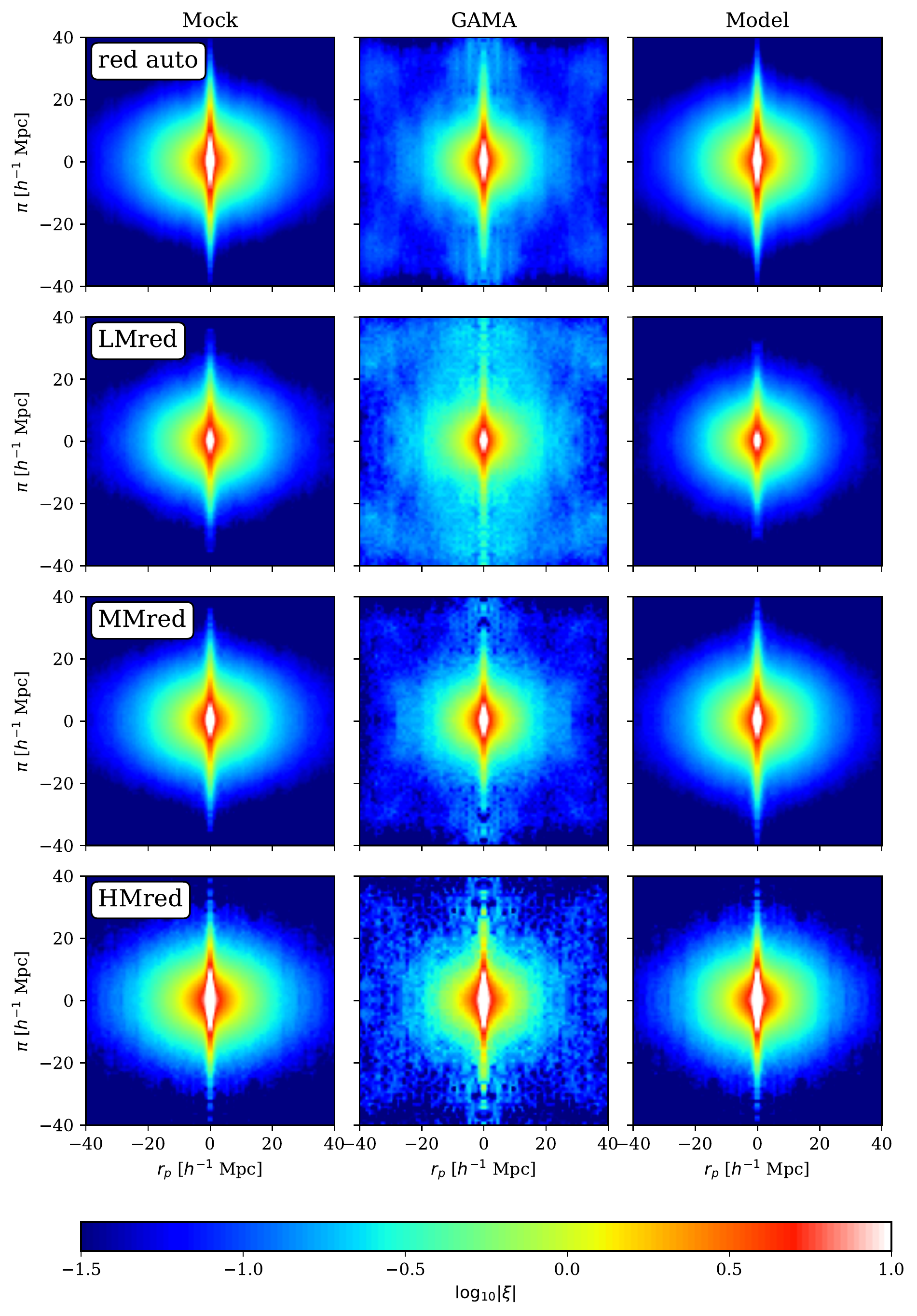}
\caption{False colour images of auto- and cross-correlation functions in redshift space for the actual red-galaxy GAMA data and the corresponding model derived according to the H (halo streaming) approach. $r_p$ denotes transverse separation; $\pi$ is radial separation. LM, MM, and HM denote the three group mass bins.}
\label{fig:gamamodred}
\end{figure*}

\begin{figure*}
\centering
\includegraphics[width=0.8\textwidth]{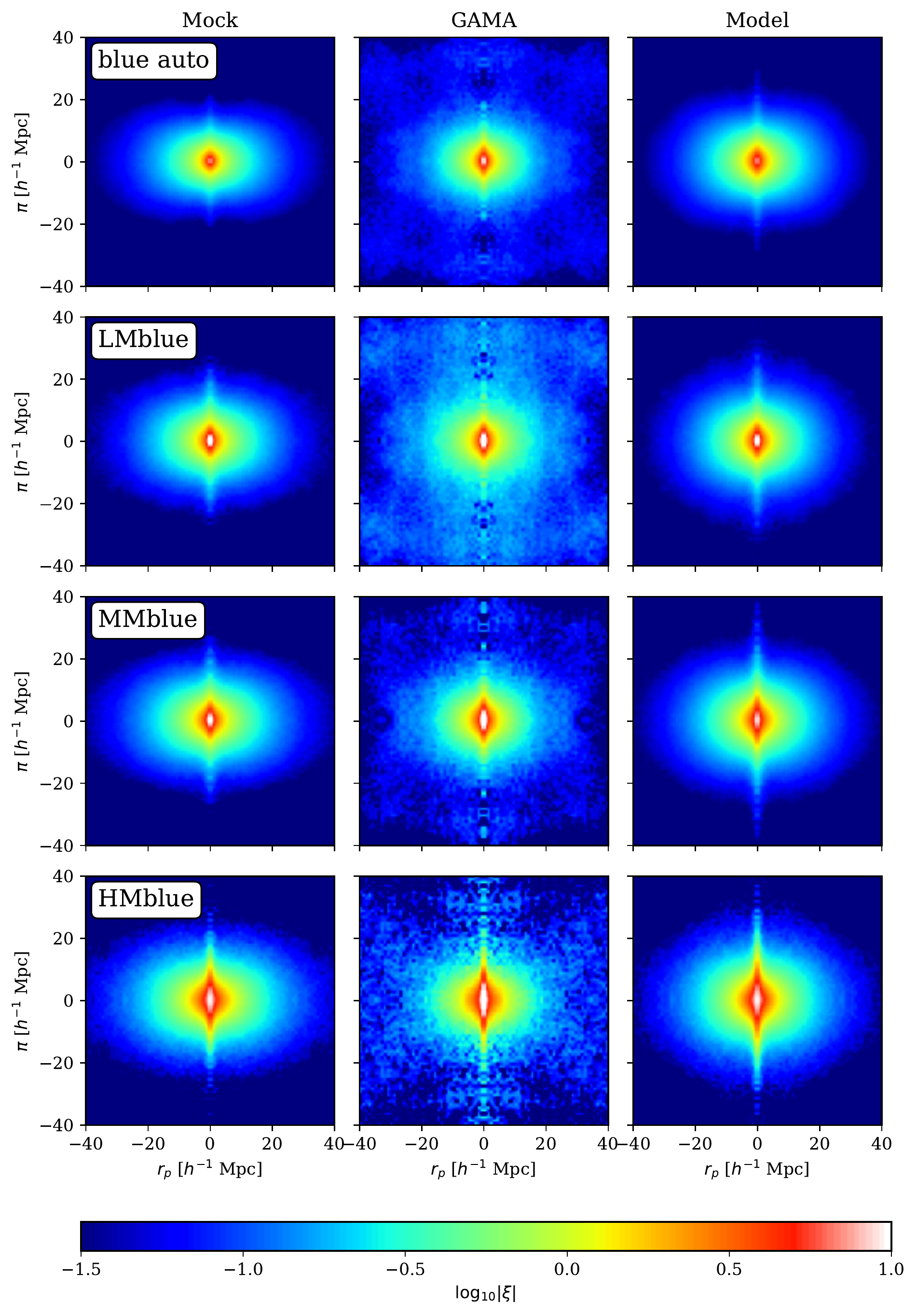}
\caption{False colour images of auto- and cross-correlation functions in redshift space for the actual blue-galaxy GAMA data and the corresponding model derived according to the H (halo streaming) approach. $r_p$ denotes transverse separation; $\pi$ is radial separation. LM, MM, and HM denote the three group mass bins.}
\label{fig:gamamodblue}
\end{figure*}

In this Appendix, we present the full 2D correlation function model fits to the GAMA data, as obtained using the HS (halo streaming) model. Fig. \ref{fig:gamamodred} shows the results for 
red galaxies and Fig. \ref{fig:gamamodblue} shows the results for blue galaxies.

\section{Summary of model parameters}
\label{appendix: model pars}

We summarise the mean and standard deviation of the model parameters for the quasilinear dispersion (QD) model (Tab.~\ref{tab: Model1 sum mock} and \ref{tab: Model1 sum data}) and the halo streaming (HS) model (Tab.~\ref{tab: Model2 sum mock} and \ref{tab: Model2 sum gama}) for the auto-correlation only and the six cross-correlation configurations.
Tab.~\ref{tab: Model1 sum mock} and \ref{tab: Model2 sum mock} show the mock results, where the mean and scatter from the best-fit of each mock sample is presented.
Tab.~\ref{tab: Model1 sum data} and \ref{tab: Model2 sum gama} show the GAMA data results, where the mean and scatter from MCMC is presented.

\begin{table*}
\centering
\caption{Mock measurements using the quasilinear dispersion (QD) model. The values are at smallest $r_{\rm min}$ which gives $f$ below $10\%$ bias compared to the fiducial value. $\sigma_8$ has been fixed to $0.81$.}
\label{tab: Model1 sum mock}
\begin{tabular}{l|c|c|c|c|c|c|c}
\hline
& All & LMred & LMblue & MMred & MMblue & HMred & HMblue\\
\hline
$r_{\rm min}$ & 10 & 15 & 10 & 20 & 10 & 20 & 10\\
$\bar{\chi}^2/{\rm dof}$ & 61.9/101 & 44.7/75 & 62.7/101 & 29.8/49 & 62.4/101 & 30.1/49 & 62.8/101\\
$f$ & 0.55 $\pm$ 0.19 & 0.56 $\pm$ 0.26 & 0.57 $\pm$ 0.16 & 0.58 $\pm$ 0.36 & 0.55 $\pm$ 0.19 & 0.63 $\pm$ 0.37 & 0.54 $\pm$ 0.18\\
$b_{\rm gal}$ & $1.22\pm0.08$ & $1.47\pm0.14$ & $0.89\pm0.06$ & $1.44\pm0.20$ & $0.90\pm0.07$ & $1.41\pm0.21$ & $0.90\pm0.06$\\\
$b_{12}$ & $0.86\pm0.02$ & $1.26\pm0.17$ & $0.74\pm0.05$ & $0.98\pm0.10$ & $0.62\pm0.03$ & $0.74\pm0.09$ & $0.43\pm0.03$\\
$\sigma_a$ & $487\pm106$ & $496\pm237$ & $392\pm107$ & $550\pm340$ & $379\pm130$ & $601\pm417$ & $378\pm117$\\
$\sigma_c$ & $473\pm109$ & $489\pm179$ & $437\pm116$ & $533\pm356$ & $410\pm120$ & $556\pm357$ & $350\pm129$\\
$I_a$ & $0.019\pm0.012$ & $0.028\pm0.020$ & $0.008\pm0.007$ & $0.029\pm0.023$ & $0.008\pm0.006$ & $0.023\pm0.022$ & $0.009\pm0.006$\\
$I_c$ & $0.021\pm0.014$ & $0.026\pm0.031$ & $0.014\pm0.020$ & $0.029\pm0.026$ & $0.015\pm0.013$ & $0.028\pm0.029$ & $0.023\pm0.016$\\
\hline
$b_{\rm grp}$ & $1.42\pm0.09$ & $1.20\pm0.22$ & $1.20\pm0.11$ & $1.48\pm0.23$ & $1.46\pm0.09$ & $1.90\pm0.30$ & $2.09\pm0.17$\\
\hline
\end{tabular}
\end{table*}

\begin{table*}
\begin{center}
\caption{Same as Table~\ref{tab: Model1 sum mock}, but for the GAMA measurements. The error bars come from MCMC posterior.}
\label{tab: Model1 sum data}
\begin{tabular}{l|c|c|c|c|c|c|c}
\hline
& All & LMred & LMblue & MMred & MMblue & HMred & HMblue\\
\hline
$r_{\rm min}$ & 10 & 15 & 10 & 20 & 10 & 20 & 10\\
$\bar{\chi}^2/{\rm dof}$ & 76/101 & 39/75 & 78/101 & 21/49 & 79/101 & 36/49 & 76/101\\
$f$ & 0.33 $\pm$ 0.15 & 0.37 $\pm$ 0.26 & 0.35 $\pm$ 0.10 & 0.37 $\pm$ 0.25 & 0.33 $\pm$ 0.10 & 0.43 $\pm$ 0.28 & 0.35 $\pm$ 0.11\\
$b_{\rm gal}$ & $1.32\pm0.08$ & $1.19\pm0.16$ & $1.03\pm0.05$ & $1.31\pm0.19$ & $1.06\pm0.05$ & $1.32\pm0.19$ & $1.04\pm0.06$\\
$b_{12}$ & $0.87\pm0.02$ & $1.11\pm0.20$ & $0.81\pm0.05$ & $0.83\pm0.11$ & $0.70\pm0.03$ & $0.77\pm0.18$ & $0.53\pm0.04$\\
$\sigma_a$ & $378\pm95$ & $670\pm250$ & $268\pm85$ & $631\pm283$ & $258\pm78$ & $595\pm276$ & $273\pm83$\\
$\sigma_c$ & $363\pm99$ & $684\pm276$ & $298\pm90$ & $622\pm252$ & $297\pm72$ & $636\pm263$ & $278\pm81$\\
$I_a$ & $0.015\pm0.005$ & $0.014\pm0.011$ & $0.005\pm0.004$ & $0.011\pm0.009$ & $0.003\pm0.003$ & $0.011+0.009$ & $0.003\pm0.003$\\
$I_c$ & $0.006 \pm 0.005$ & $0.010\pm0.009$ & $0.007\pm0.006$ & $0.036\pm0.018$ & $0.026\pm0.025$ & $0.038\pm0.023$ & $0.037\pm0.016$\\
\hline
$b_{\rm grp}$ & $1.52\pm0.10$ & $1.07\pm0.24$ & $1.27\pm0.10$ & $1.58\pm0.30$ & $1.52\pm0.10$ & $1.71\pm0.46$ & $1.98\pm0.19$\\
\hline
\end{tabular}
\end{center}
\end{table*}

\begin{table*}
\centering
\caption{Mock measurements using the halo streaming (HS) model. $\sigma_8$ has been fixed to $0.81$. The error bars come from the standard deviation of best-fit values over the 25 mock samples.}
\label{tab: Model2 sum mock}
\begin{tabular}{l|c|c|c|c|c|c}
\hline
& LMred & LMblue & MMred & MMblue & HMred & HMblue\\
\hline
$r_{\rm min}$ & 5 & 5 & 5 & 5 & 5 & 5 \\
$\bar{\chi}^2/{\rm dof}$ & $77\pm15$/118 & $77\pm11$/118 & $81\pm12$/118 & $78\pm11$/118 & $81\pm15$/118 & $76\pm11$/118\\
$f$ & $0.61\pm0.17$ & $0.60\pm0.13$ & $0.57\pm0.19$ & $0.62\pm0.16$ & $0.59\pm0.16$ & $0.62\pm0.15$\\
$b_{\rm gal}$ & $1.38\pm0.08$ & $0.87\pm0.04$ & $1.38\pm0.07$ & $0.87\pm0.05$ & $1.37\pm0.07$ & $0.87\pm0.05$\\\
$b_{12}$ & $1.17\pm0.09$ & $0.73\pm0.06$ & $0.97\pm0.04$ & $0.63\pm0.04$ & $0.70\pm0.06$ & $0.46\pm0.04$\\
$\sigma_a$ & $482\pm76$ & $358\pm67$ & $459\pm95$ & $369\pm84$ & $467\pm71$ & $371\pm66$\\
$\sigma_c$ & $524\pm87$ & $458\pm77$ & $484\pm99$ & $417\pm94$ & $469\pm81$ & $383\pm85$\\

$\alpha_a$ & $0.99\pm0.08$ & $1.00\pm0.18$ & $0.99\pm0.11$ & $1.04\pm0.22$ & $0.98\pm0.11$ & $1.02\pm0.23$\\
$\alpha_c$ & $0.96\pm0.09$ & $0.95\pm0.19$ & $0.96\pm0.10$ & $1.00\pm0.13$ & $1.01\pm0.08$ & $1.02\pm0.12$\\
$\eta_a$ & $1.02\pm0.05$ & $1.01\pm0.04$ & $1.00\pm0.05$ & $1.00\pm0.03$ & $1.01\pm0.06$ & $0.99\pm0.05$\\
$\eta_c$ & $1.04\pm0.04$ & $1.02\pm0.04$ & $1.02\pm0.04$ & $1.00\pm0.04$ & $0.99\pm0.04$ & $0.98\pm0.05$\\
$I_a$ & $0.0035$ & $0.0015$ & $0.0035$ & $0.0015$ & $0.0035$ & $0.0015$\\
$I_c$ & $0.0020$ & $0.0018$ & $0.0032$ & $0.0018$ & $0.0036$ & $0.0019$\\
\hline
$b_{\rm grp}$ & $1.18\pm0.13$ & $1.20\pm0.12$ & $1.42\pm0.09$ & $1.38\pm0.09$ & $1.96\pm0.17$ & $1.92\pm0.16$\\
\hline
\end{tabular}
\end{table*}

\begin{table*}
\centering
\caption{Same as Tab.~\ref{tab: Model2 sum mock} but for the GAMA data. The error bars come from MCMC posterior.}
\label{tab: Model2 sum gama}
\begin{tabular}{l|c|c|c|c|c|c|c|c}
\hline
& red & blue & LMred & LMblue & MMred & MMblue & HMred & HMblue\\
\hline
$r_{\rm min}$ & 5 & 5 & 5 & 5 & 5 & 5 & 5 & 5 \\
$\bar{\chi}^2/{\rm dof}$ & $40$/59 & $41$/59 & $68$/118 & $88$/118 & $80$/118 & $89$/118 & $85$/118 & $84$/118\\

$f$ & $0.49\pm0.24$ & $0.39\pm0.11$ & $0.40\pm0.20$ & $0.33\pm0.08$ & $0.41\pm0.20$ & $0.39\pm0.11$ & $0.45\pm0.22$ & $0.37\pm0.11$\\

$b_{\rm gal}$ & $1.34\pm0.09$ & $1.00\pm0.04$ & $1.26\pm0.08$ & $1.01\pm0.04$ & $1.40\pm0.08$ & $1.02\pm0.04$ & $1.36\pm0.09$ & $1.02\pm0.04$\\

$b_{12}$ & - & - & $1.10\pm0.08$ & $0.81\pm0.04$ & $0.99\pm0.04$ & $0.76\pm0.03$ & $0.75\pm0.07$ & $0.57\pm0.04$\\

$\sigma_a$ & $454\pm130$ & $242\pm73$ & $350\pm109$ & $197\pm60$ & $400\pm101$ & $270\pm77$ & $459\pm121$ & $247\pm80$\\

$\sigma_c$ & - & - & $391\pm122$ & $325\pm67$ & $440\pm99$ & $281\pm69$ & $378\pm145$ & $263\pm83$\\

$\alpha_a$ & $1.07\pm0.14$ & $0.82\pm0.26$ & $0.95\pm0.11$ & $0.70\pm0.18$ & $0.89\pm0.09$ & $0.83\pm0.23$ & $1.05\pm0.14$ & $0.78\pm0.28$\\

$\alpha_c$ & - & - & $0.88\pm0.20$ & $0.80\pm0.29$ & $0.93\pm0.14$ & $0.69\pm0.15$ & $1.04\pm0.22$ & $0.94\pm0.18$\\

$\eta_a$ & $0.76\pm0.14$ & $1.54\pm0.21$ & $0.77\pm0.12$ & $1.67\pm0.21$ & $1.03\pm0.14$ & $1.51\pm0.20$ & $0.79\pm0.14$ & $1.56\pm0.24$\\

$\eta_c$ & - & - & $0.87\pm0.16$ & $1.21\pm0.21$ & $1.11\pm0.13$ & $1.62\pm0.19$ & $0.99\pm0.13$ & $1.32\pm0.17$\\

$I_a$ & $0.0035$ & $0.0015$ & $0.0035$ & $0.0015$ & $0.0035$ & $0.0015$ & $0.0035$ & $0.0015$\\
$I_c$ & - & - & $0.0020$ & $0.0018$ & $0.0032$ & $0.0018$ & $0.0036$ & $0.0019$\\
\hline
$b_{\rm grp}$ & - & - & $1.14\pm0.11$ & $1.24\pm0.08$ & $1.41\pm0.10$ & $1.34\pm0.07$ & $1.81\pm0.20$ & $1.79\pm0.14$\\
\hline
\end{tabular}
\end{table*}

\begin{table*}
\centering
\caption{Same as Tab.~\ref{tab: Model2 sum gama} but for the 1-halo templated adopted from the mock contaminated red samples.}
\label{tab: Model2 sum gama cont red}
\begin{tabular}{l|c|c|c}
\hline
& LMred$^*$ & MMred$^*$ & HMred$^*$\\
\hline
$r_{\rm min}$ & 5 & 5 & 5 \\
$\bar{\chi}^2/{\rm dof}$ & $119$/118 & $137$/118 & $144$/118 \\
$f$ & $0.52\pm0.17$ & $0.48\pm0.17$ & $0.42\pm0.16$ \\
$b_{\rm gal}$ & $1.24\pm0.07$ & $1.35\pm0.06$ & $1.33\pm0.06$ \\
$b_{12}$ & $1.17\pm0.10$ & $0.95\pm0.03$ & $0.74\pm0.05$ \\
$\sigma_a$ & $424\pm85$ & $427\pm76$ & $427\pm93$ \\
$\sigma_c$ & $454\pm114$ & $492\pm75$ & $362\pm116$ \\

$\alpha_a$ & $1.51\pm0.16$ & $1.37\pm0.10$ & $1.51\pm0.16$ \\
$\alpha_c$ & $1.16\pm0.21$ & $1.14\pm0.11$ & $1.22\pm0.18$ \\
$\eta_a$ & $0.81\pm0.10$ & $1.03\pm0.10$ & $0.90\pm0.12$ \\
$\eta_c$ & $0.86\pm0.15$ & $1.12\pm0.07$ & $1.04\pm0.10$ \\
$I_a$ & $0.0031$ & $0.0031$ & $0.0031$ \\
$I_c$ & $0.0011$ & $0.0046$ & $0.0049$ \\
\hline
$b_{\rm grp}$ & $1.05\pm0.11$ & $1.42\pm0.08$ & $1.78\pm0.15$\\
\hline
\end{tabular}
\end{table*}

\section{MCMC posteriors}
\label{appendix: mcmc}

\begin{figure*}
\centering
    \includegraphics[width=\textwidth]{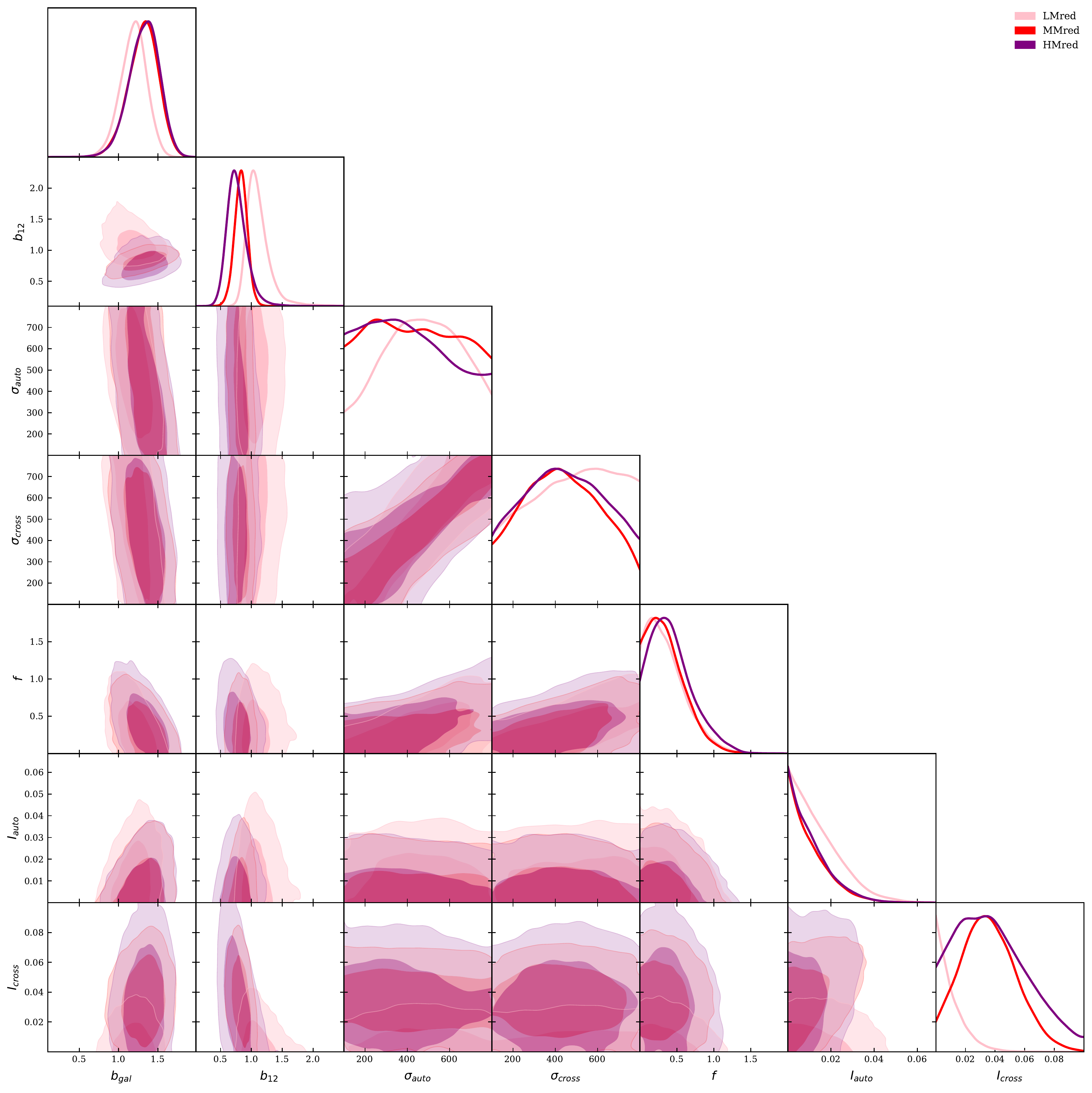}
    \caption{The posteriors for fitting the QD model to the red galaxy cross-correlations.}
    \label{fig: M1-autocolor-GAMA-blue-MCMC}
\end{figure*}

\begin{figure*}
\centering
    \includegraphics[width=\textwidth]{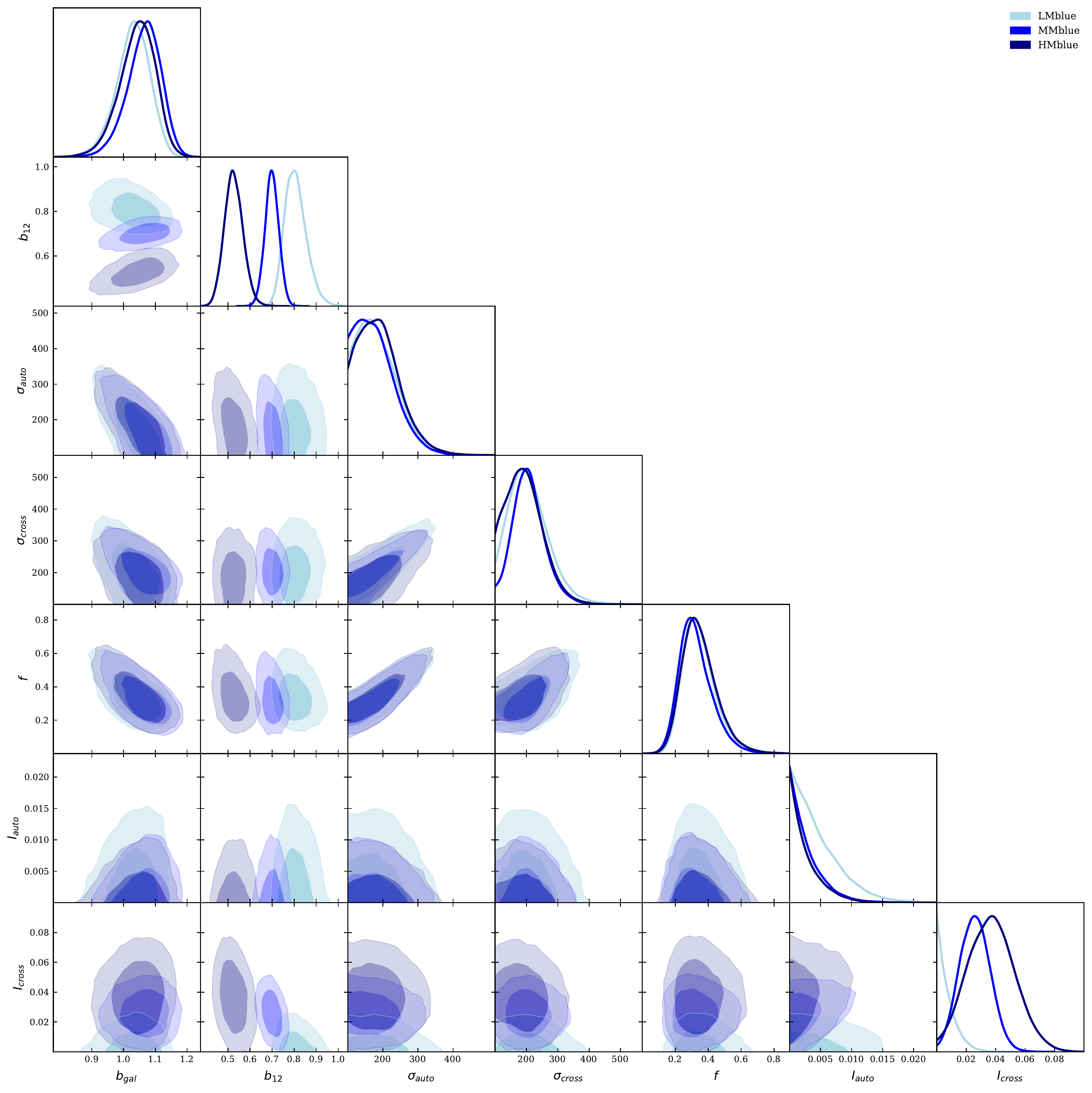}
    \caption{The posteriors for fitting the QD model to the blue galaxy cross-correlations.}
    \label{fig: M1-autocolor-GAMA-red-MCMC}
\end{figure*}

\begin{figure*}
\centering
    \includegraphics[width=\textwidth]{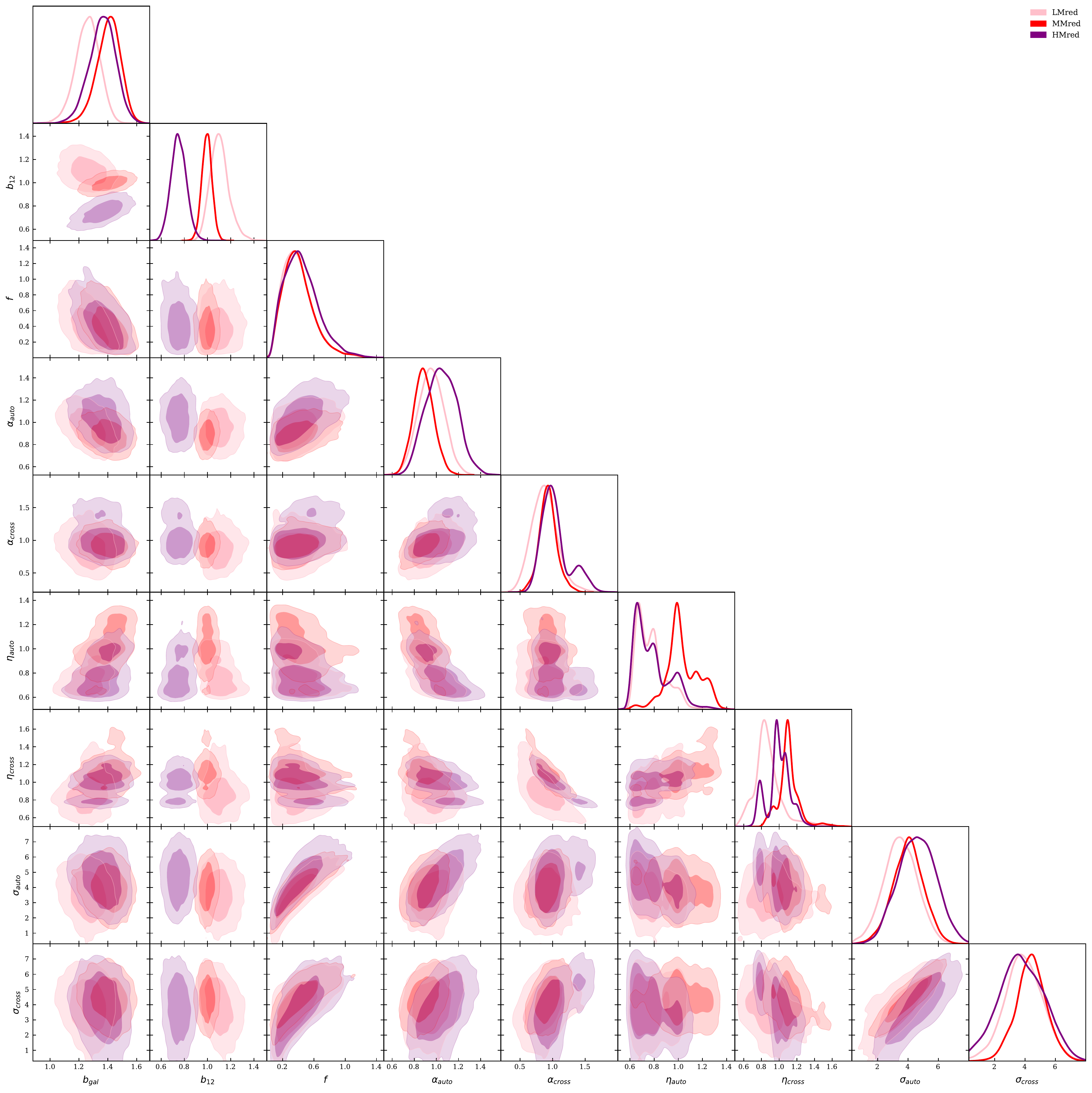}
    \caption{The posteriors for fitting the HS model to the red galaxy cross-correlations.}
    \label{fig: GSM-autocolor-GAMA-red-MCMC}
\end{figure*}

\begin{figure*}
\centering
    \includegraphics[width=\textwidth]{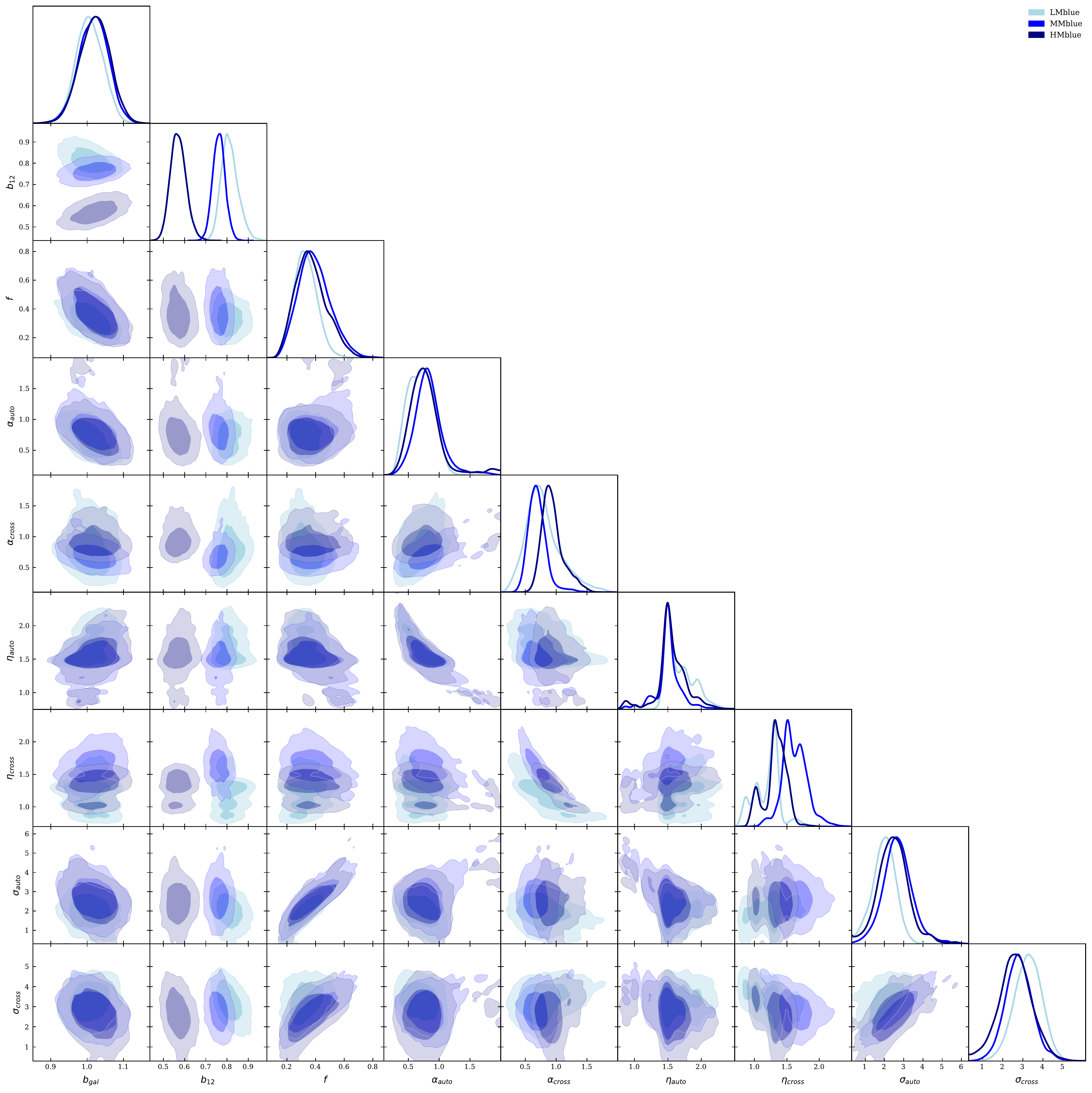}
    \caption{The posteriors for fitting the HS model to the blue galaxy cross-correlations.}
    \label{fig: GSM-autocolor-GAMA-blue-MCMC}
\end{figure*}

We present the full parameter posterior for the QD and HS model applied to the GAMA measurements for the six configurations in this section. Fig.~\ref{fig: M1-autocolor-GAMA-blue-MCMC} shows the three red configurations in the QD model, and Fig.~\ref{fig: M1-autocolor-GAMA-red-MCMC} shows the blue configurations. Fig.~\ref{fig: GSM-autocolor-GAMA-red-MCMC} shows the three red configurations in the HS model, and Fig.~\ref{fig: GSM-autocolor-GAMA-blue-MCMC}.

\bsp	
\label{lastpage}
\end{document}